\documentclass[12pt]{article}
\usepackage[utf8]{inputenc}
\usepackage[T1]{fontenc}
\usepackage{amsmath}
\usepackage{amsthm}
\usepackage{amssymb}
\usepackage{amsfonts}
\usepackage{times}
\usepackage{fancyhdr}
\usepackage[ruled,vlined]{algorithm2e}
\usepackage{mathrsfs}
\usepackage{stmaryrd}
\usepackage{tabularx}
\usepackage{threeparttable}
\usepackage[export]{adjustbox}
\usepackage{graphicx,psfrag,epsf}
\usepackage{enumerate}
\usepackage{smile}
\usepackage{appendix}
\usepackage{tocbasic}
\usepackage{setspace}
\usepackage[margin=2cm]{caption}
\usepackage{natbib}
\usepackage{multirow}
\usepackage{lscape}
\usepackage{subfigure}
\usepackage{makecell}
\usepackage{exscale}
\usepackage{booktabs}
\usepackage{xr}
\usepackage{array}
\usepackage{url}
\usepackage{lineno}
\usepackage{algpseudocode}
\usepackage{float}
\usepackage{threeparttable}
\usepackage{bm}
\usepackage{mathtools}
\usepackage{wrapfig}
\usepackage{lipsum}
\usepackage{mathrsfs}
\usepackage{multirow}
\graphicspath{ {./images/} }
\usepackage{apalike}
\usepackage[usenames,dvipsnames,svgnames,table]{xcolor}
\usepackage[colorlinks=true,
linkcolor=blue,
urlcolor=blue,
citecolor=blue]{hyperref}
\usepackage{url}
\usepackage{xr}
\numberwithin{equation}{section}
\numberwithin{theorem}{section}
\numberwithin{corollary}{section}
\numberwithin{definition}{section}

\newtheorem{thm1}{Theorem}[section]

\newtheorem{as1}{Assumption}
\newtheorem{rmk1}{Remark}[section]

\begin{document}
		\title{\bf Factor Modelling for Biclustering Large-dimensional Matrix-valued Time Series}
	\author{Yong He\thanks{ Institute for Financial Studies, Shandong University, Jinan, China; Email:{\tt heyong@sdu.edu.cn}.},~~Xiaoyang Ma\thanks{Institute for Financial Studies, Shandong University, Jinan, China; Email:{\tt maxiaoyang@mail.sdu.edu.cn}.},~~Xingheng Wang\thanks{ Fudan University, Shanghai, China; Email:{\tt xinghengw1@gmail.com}.},~~Yalin Wang\thanks{School of Mathematics, Shandong University, Jinan, China; Email:{\tt 
wangyalin@mail.sdu.edu.cn }.}}	

	\date{{\small \today}}
	\maketitle 

	\begin{abstract}
    A novel unsupervised learning method is proposed in this paper for biclustering large-dimensional matrix-valued time series based on an entirely new latent two-way factor structure. Each block cluster is characterized by its own row and column cluster-specific factors in addition to some common matrix factors which impact on all the matrix time series. We  first estimate the global loading spaces by projecting  the observation matrices onto
the row or column loading space corresponding to common factors. The loading spaces for cluster-specific factors are then further recovered by projecting  the observation matrices onto the orthogonal complement space of the estimated global loading spaces. 
    To identify the latent row/column clusters simultaneously for matrix-valued time series, we provide a $K$-means algorithm based on the estimated row/column factor loadings of the cluster-specific weak factors. Theoretically, we derive faster convergence rates for global loading matrices than those of the state-of-the-art methods available in the literature under mild conditions. We also propose an one-pass eigenvalue-ratio method to estimate the numbers of global and cluster-specific factors. The consistency with explicit convergence rates is also established for the estimators of the local loading matrices, the factor numbers and the latent cluster memberships.
    Numerical experiments with both simulated data as well as a real data example are also reported to illustrate the usefulness of our proposed method.
\end{abstract}
	
	\vfill
	
	\newpage

\section{Introduction}\label{sec1}
\begin{figure}[ht!]
    \centering
\includegraphics[width=16cm,height=8cm]{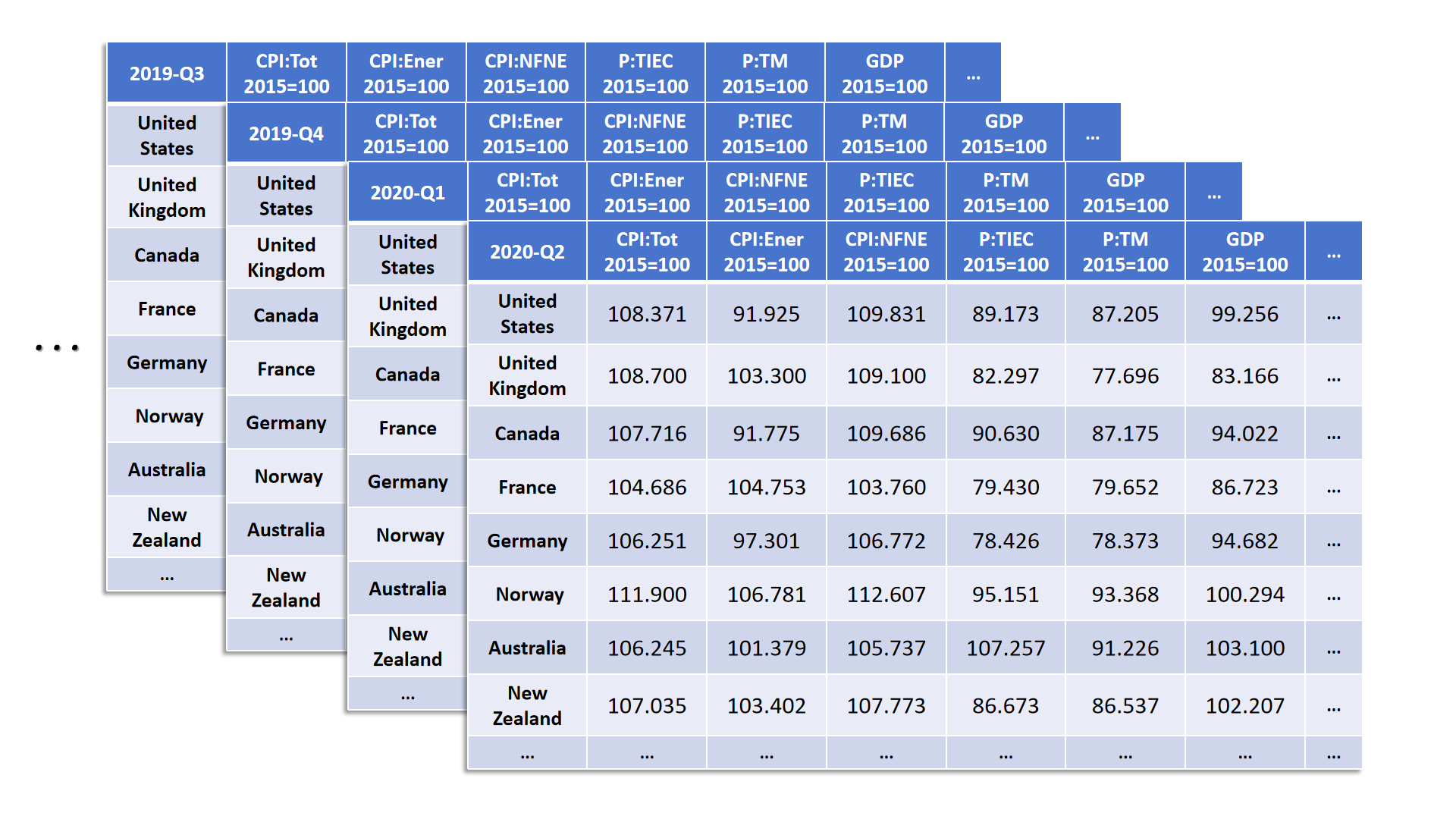}
\caption{A real example of matrix-variate observations consisting of macroeconomic variables across multiple countries.}
    \label{macroeconomic_data}
\end{figure}

Matrix-valued time series data are prevalent in various research fields such as economics, finance, and engineering, and have been extensively studied in the statistician and econometrician communities \citep{Samadi2014MatrixTS,Ding2018MatrixVR,Wang2019,Chen2021AutoregressiveMF,chen2022factor,YU2022,Fan2023,He2024Matrix}. From financial markets to Industrial Manufacturing, large number of time series typically have latent group structures and unsupervised clustering analysis for time series data has gained growing attention. For better illustration, Figure \ref{macroeconomic_data} shows a time list of tables recording the macroeconomic variables across a number of worldwide countries. As a common sense, the column indicators can be roughly divided into several categories, namely Consumer Price, Interest Rate, Production, and International Trade, meanwhile  the countries can also be grouped according to their  geographic location or economic development level, namely developed, developing and 
underdeveloped countries. Clustering large-scale time series data into latent groups would help better understand the underlying mechanism  of data generation, thereby beneficial for future prediction. 
 However, as far as we know, clustering matrix-valued time series has barely been discussed in the literature due to the challenge brought by the interactions across both rows and columns.  The main focus of the current work is to address this challenging problem with an entirely new matrix factor model setup.
In the following we briefly review the closely related literature with our work.

\subsection{Closely Related Work}
Factor models have been widely used in the field of finance, economics, and various other disciplines \citep{stock2002b,bai2002determining,bai2003inferential,onatski2009testing,fan2013large,  trapani2018randomized,He2022LargeDimensionalFA}, serving as an important dimension reduction tool to analyze high-dimensional datasets by characterizing the dependency structure of variables via a few latent factors. The time list of tables recording the macroeconomic variables across a number of worldwide countries shown in Figure \ref{macroeconomic_data} is an typical example of matrix-valued time series. A possible approach to analyze matrix-valued time series is to first vectorize the data and then employ the techniques developed for vector time series, but this would lead to suboptimal
inference \citep{YU2022,Fan2023,he2024online}. 
 \cite{Wang2019} was the first to introduce the following factor model for matrix time series $\{\mathbf{X}_t,1\leq t\leq T\}$:
\begin{equation}\label{factor_model}
    \left(\mathbf{X}_t\right)_{p \times q}=(\mathbf{R})_{p \times k}\left(\mathbf{F}_t\right)_{k \times r}\left(\mathbf{C}^{\top}\right)_{r \times q}+\left(\mathbf{E}_t\right)_{p \times q}, \quad t=1, \ldots, T
\end{equation}
where $\mathbf{R}$ and $\mathbf{C}$ are respectively the  row and column factor loading matrices, capturing the variations in $\mathbf{X}_t$ across the rows and columns. The lower-dimensional matrices $\mathbf{F}_t$ represent the common factors influencing all elements of $\mathbf{X}_t$, while $\mathbf{E}_t$ denote the idiosyncratic components that be viewed as vector white noises in \cite{Wang2019}. Model (\ref{factor_model}) has drawn growing attention in the last few years and further been extended to accommodate more complex scenarios.  For instance, \cite{Liu2019HelpingEA} introduce a threshold factor models to analyze matrix-valued time series; \cite{Chen2020ConstrainedFM} establish a general framework for incorporating domain  knowledge in the matrix factor model through linear constraints; \cite{YU2022} propose a projected estimation approach, which achieves faster convergence rates by increasing signal-to-noise ratio. Model (\ref{factor_model}) along with its variants have also
been studied by, but are not limited to, \cite{Han2021Rank,chen2022factor,Kong2022MatrixQF,He2023Vector,He2024Matrix}.

Existing methods for analyzing matrix factor models  predominantly rely on the assumption of strong factors. However, in practical applications, the strong factor assumption may no longer hold due to the presence of missing data or underlying group structures.  In fact, most existing work on factor modeling  assume all factors are strong,  which results in a clear partition of the
eigenvalues of the observed covariance matrix into two sets: large eigenvalues representing factor-related
variation and small eigenvalues representing idiosyncratic variation.  However, empirical studies in economics and finance indicate
that eigenvalues often diverge at varying rates  and influential empirical studies especially on asset pricing often give implicit yet strong evidence of weak factor models \citep{uematsu2022estimation,massacci2024instability}. A few works discuss the possible existence of weak factors for matrix factor model, such as \cite{Wang2019,Chen2020ConstrainedFM,He2023Vector}, which all consider weak factors along both the row and column dimensions and model their strength  similar to Assumption \ref{ass1} below.

Another closely related strand of research focuses on clustering for large-dimensional time series. Factor models with latent group structures are a relatively new topic in econometrics \citep{bonhomme2015grouped}. To the best of our knowledge, only a few works propose clustering analysis methods in the context of large-dimensional factor models. Inspired by the  popping up research of panel regression models with group structure of the regression coefficients \citep{vogt2017classification,wang2018homogeneity,chen2019estimating,chen2021nonparametric}, \cite{tu2023}  propose an unsupervised clustering method, which identifies the grouping structure of factor loadings in the large-dimensional approximate factor model; \cite{Yao2023} propose a new method for clustering a large number of time series based on a latent factor structure and represent the dynamic structures by latent common and cluster-specific factors. \cite{he2024penalized}  propose a fusion Penalized Principal Component Analysis (PPCA) method to  identify group structures within the framework of large-dimensional approximate factor models. Existing clustering methods on factor models mainly focus on vector time series and it is still vacant for matrix-valued time series to the best of our knowledge, which is quite challenging due to the two-way interactions across both
rows and columns. A closely related concept is biclustering, also called block clustering, co-clustering, two-way clustering or two-mode clustering, which clusters the rows and columns in a matrix simultaneously.  Biclustering was initially proposed by \cite{Hartigan1972DirectCO}, becoming growing important in biology and biomedicine for gene expression analysis with the advent of the \cite{Cheng2000BiclusteringOE}'s algorithm. Recently,  biclustering has been shown of great use in domains such as text mining, recommendation systems, and climate science \citep{Singh2018ScalabilityAS,Wu2020AnOO}.  
\subsection{Contributions}
In this article, we propose a new model setup for biclustering matrix-valued time series, in which we capture the dynamic structures of matrix-valued time series by common factors and cluster-specific factors that are all latent and can be separated by their factor strengths. Our goal and methodology are radically different from those of the vast literature on matrix factor model. The goal of this study is to recover the latent row/column clusters simultaneously for matrix-valued time series by a matrix factor model with both strong and weak factors, where the weak factors correspond to the cluster-specific dynamics. We first estimate the factor loading spaces for common factors  by projecting  the observation matrices onto
the row or column loading space corresponding to common factors. We then further estimate the loading matrices for cluster-specific factors by projecting the observation matrices
onto the orthogonal complement space of the estimated global loading space. We also propose an eigenvalue-ratio  method to estimate the numbers of common and cluster-specific factors. To identify the memberships of row/column clusters simultaneously for matrix-valued time series, we provide a $K$-means algorithm based on the estimated row/column factor loadings of the cluster-
specific weak factors. In summary, the contributions of the current work lie in the following aspects: 
firstly, the proposed biclustering algorithm based on matrix factor model serves as a much-needed addition to the vacant literature on clustering analysis for large-dimensional matrix time series and the projection technique to estimate both the strong and weak factors is of independent interest to the related literature.
Secondly, our model setup is quite general in the sense that we allow dependence between the common factors and cluster-specific factors and our  method remains feasible even if the idiosyncratic components exhibit weak serial correlations, which is usually assumed to be white noise in the literature. 
Thirdly, we focus on  dynamically dependent factors and propose an auto-covariance based method to estimate the factor loading matrices inspired by \cite{Lam2012FactorMF}. The convergence rate of the estimated global loading matrices attains $T^{-1/2}$ in terms of the averaged $L_2$-norm, which is much faster than the rate obtained by \cite{Wang2019}. We also derive the convergence rates of the estimated cluster-specific loading matrices, which is particularly challenging due to the weak serial correlation in the noise, the presence of both strong and weak factors, and the interaction terms in the signal part of the matrix factor model. 
Finally, we utilize an eigenvalue-ratio based, one-pass method for estimating the numbers of strong and weak factors across both the row and column dimensions. The eigenvalue-ratio based estimators are shown to be consistent, which serves as another valuable theoretical and methodological contribution.

\subsection{Organizations and Notations}
The rest of this paper is organized as follows. In Section \ref{sec2}, we introduce the model setup and illustrate the procedure to estimate the loadings and identify the underlying cluster structure. In Section \ref{sec3}, we investigate the theoretical properties of the proposed estimators of loading matrices, factor numbers and cluster memberships. Simulation results are shown in Section \ref{sec4}. We also use the proposed method to analyze a large-scale real macroeconomic dataset in Section \ref{sec5}. Further discussions and future research directions are left in Section \ref{sec6}. Proofs of the main theorems and technical lemmas are delegated to the Supplementary Material.

To end this section, we introduce some notations used throughout the study. For any matrix-valued time series $\mathbf{X}_t$, let the $i$-th row of $\mathbf{X}_t$ be $\mathbf{x}_{t,i.}$, both $\mathbf{x}_{t,.j}$ and $\mathbf{x}_{t,j}$ denote the $j$-th column of $\mathbf{X}_t$. For any vector $\boldsymbol{a}$ and $\boldsymbol{b}$,  $\|\boldsymbol{a}\|_q$ denotes its $L_q$-norm,  $q\ge 1$, $\|\boldsymbol{a}\|$ also denotes $L_2$-norm. $\boldsymbol{a}\stackrel{d}{=}\boldsymbol{b}$ means distributions of $\boldsymbol{a}$ and $\boldsymbol{b}$ are the same. For a (random) matrix $\mathbf{A}=(a_{ij})_{p \times q}$ of dimension \( p \times q \), $\mathcal{M}(\mathbf{A})$ denotes the linear space spanned by the columns of $\mathbf{A}$. $|\mathbf{A}|$ denotes the matrix with $|a_{ij}|$ as its $(i,j)$-th element. We use $\mathbf{A}'$ and $\mathbf{A}^{\top}$ to represent the transpose of matrix $\mathbf{A}$, both $\|\mathbf{A}\|_2$ and $\|\mathbf{A}\|$ denote the spectral norm of  $\mathbf{A} $, $\|\mathbf{A}\|_{\max}=\max_{i,j}|a_{ij}|$ is the max norm of $\mathbf{A}$, $\|\mathbf{A}\|_{\min}$ denotes the square root of the minimum non-zero eigenvalue of $\mathbf{A}^{\top}\mathbf{A}$ and $\|\mathbf{A}\|_F$ denotes the Frobenius norm of $\mathbf{A}$. Let $\text{tr}(\mathbf{A})$ be the trace of $\mathbf{A}$. Define ${\mathbf{P}}_{{A}}:=\mathbf{A}(\mathbf{A}^{\top}\mathbf{A})^{-1}\mathbf{A}^{\top}$, thus ${\mathbf{P}}_{{A}}$ denotes the projection matrix of the linear space spanned by the columns of matrix $\mathbf{A}$ and ${\mathbf{P}}_{{A}}^c=\mathbf{I}_p-{\mathbf{P}}_{{A}}$ is the projection matrix of its orthogonal complement space. Denote $\mathbf{A}_{i \cdot}$ and $\mathbf{A}_{\cdot j}$ respectively as the $i$-th row and $j$-th column of $\mathbf{A}$. $\mathbf{I}_q$ denotes the identity matrix of order $q$. Let $\lambda_i(\mathbf{A})$ be the $i$-th eigenvalue of the non-negative definite matrix $\mathbf{A}$ in descending order. For two series $\left\{a_n\right\}_{n \geq 1}$ and $\left\{b_n\right\}_{n \geq 1}$, if there is a constant $C$ such that $a_n \geq C b_n \ (a_n \leq C b_n), \forall n \geq 0$, we denote it as $a_n \gtrsim b_n \ (a_n \lesssim b_n)$. We write $a_n \asymp b_n$ if $a_n \gtrsim b_n$ and $a_n \lesssim b_n$ both hold. For two series of random variables $\left\{X_n\right\}_{n \geq 1}$ and $\left\{Y_n\right\}_{n \geq 1}$,  $X_n \gtrsim Y_n \ (X_n \lesssim Y_n)$ means $Y_n = O_p (X_n) 
 \ (X_n =O_p(Y_n))$, we say $X_n \asymp Y_n$ if $Y_n = O_p (X_n) $ and $X_n =O_p(Y_n)$ hold simultaneously. The constant $C$ in different lines can be nonidentical.

\section{Model Setup and Methodology}\label{sec2}
We first introduce our model setup for biclustering matrix-valued time series. For matrix-valued time series \(\mathbf{X}_t\in \mathbb{R}^{p\times q},t=1,\dots,T\), suppose that the row variables and column variables can be grouped into $m$ clusters and $n$ clusters, respectively. In other words, suppose that $\mathbf{X}_t$ consists of $mn$ latent blocks, i.e., 
\begin{equation}\label{X}
    \mathbf{X}_t=\begin{pmatrix}
    \mathbf{X}_{t,11}& \cdots & \mathbf{X}_{t,1n}\\
    \vdots & \ddots & \vdots\\
    \mathbf{X}_{t,m1}& \cdots & \mathbf{X}_{t,mn}
\end{pmatrix},
\end{equation}
where $\mathbf{X}_{t,11},\cdots,\mathbf{X}_{t,mn}$ are, respectively, $p_1 \times q_1, \cdots, p_m \times q_n$-matrix-valued time series with $p_1, \cdots,p_m \ge 1$, $q_1, \cdots,q_n \ge 1$, and $
p_1+\dots+p_m=p,\ \ \ \ q_1+\dots+q_n=q.
$

We  assume that there exists an underlying two-way factor structure in $\mathbf{X}_{t}$ and the common components consist of two parts.  The first part is led  by some global/common two-way factors, which are strong factors and affect the vast majority (if not all) of time series. The second part is driven by some cluster-specific factors, which are weak factors and only affect the time series in a specific cluster.  
In detail, we model the matrix time series $\mathbf{X}_{t}$ as 
\begin{equation}\label{model}
\mathbf{X}_t=\mathbf{R}\mathbf{G}_t\mathbf{C}^{\top}+\mathbf{\Gamma}\mathbf{F}_t{\mathbf{\Lambda}}^{\top}+\mathbf{E}_t^0.
\end{equation}
In model (\ref{model}),  \(\mathbf{R}\) is a \(p\times k_0\) global row factor loading matrix, \(\mathbf{C}\) is a \(q\times r_0\) global column factor loading matrix, \(\mathbf{G}_t\in \mathbb{R}^{k_0\times r_0}\) is the latent matrix-valued common/global factors,
 \(\mathbf{F}_t\in \mathbb{R}^{k\times r}\) contains the cluster-specific factors,  \(\mathbf{\Gamma}\) and \(\mathbf{\Lambda}\) are  the corresponding left and right cluster-specific factor loading matrices with rank \(k\) and \(r\) respectively. Both of the cluster-specific factor loading matrices \(\mathbf{\Gamma}\) and \(\mathbf{\Lambda}\) have a block diagonal structure, i.e., $\mathbf{\Gamma}=\text{diag}\left(
\mathbf{\Gamma}_1,\mathbf{\Gamma}_2,\dots,\mathbf{\Gamma}_m\right),
\mathbf{\Lambda}=\text{diag}\left(
\mathbf{\Lambda}_1,\mathbf{\Lambda}_2,\dots,\mathbf{\Lambda}_n\right)$, where \(\mathbf{\Gamma}_i\) and \(\mathbf{\Lambda}_j\) are  \(p_i\times k_i\) and \(q_j\times r_j\) matrices respectively. Thus, $k$ and $r$ represent the total number of row and column cluster-specific factors and we have
$k_1+\dots+k_m=k$ and $r_1+\dots+r_n=r$. In addition, \(\mathbf{E}_t^0\) is the idiosyncratic components. From the model setup, we see that after blocking, the elements within the same cluster in $\mathbf{X}_{t}$ are influenced not only by a subset of common factors $\mathbf{G}_t$  but also by cluster-specific factors in $\mathbf{F}_t$, i.e.,
 each sub-matrix $\mathbf{X}_{t,ij}$ is driven by the cluster-specific factor $\mathbf{F}_{t,ij}$ and the common factors $\mathbf{G}_t$. Indeed, the latent clusters are characterized by the block diagonal structures of  \(\mathbf{\Gamma}\) and \(\mathbf{\Lambda}\) and  we only observe permuted $\mathbf{X}_t$ and both row and column  cluster memberships  of $\mathbf{X}_t$ are unknown. Our goal is to cluster matrix time series into different
and unknown block-clusters, while the number of row/column clusters ($m,n$)
and their sizes $\{(k_i,r_j),i=1,\ldots,m,j=1,\ldots,n\}$ are all unknown and the terms on the RHS of (\ref{model}) are unobservable.

We always assume that both the row and column numbers of the common factors and cluster-specific factors  remain bounded when the dimensions $p$ and $q$ diverges to infinity. This reflects the fact that the factor models are only attractive when the numbers of factors are much smaller than the number of time series concerned. Furthermore, we assume that the number of time series $p_i$ ($q_i$) in each cluster diverges at a lower rate than $p$ ($q$), and the numbers of clusters $m$ and $n$ diverge as well, as explicitly stated in Assumption \ref{ass1}.

\begin{as1}\label{ass1}
$\max_{0 \leq i \leq m}\left\{k_i\right\}<C<\infty$, $\max_{0 \leq j \leq n}\left\{r_j\right\}<C<\infty$, $k \asymp m=O\left(p^{\delta_{1}}\right)$, $r \asymp n=O\left(q^{\delta_{2}}\right)$, $p_i \asymp p^{1-\delta_{1}}$ for $i=1, \cdots, m$, and $q_j \asymp q^{1-\delta_{2}}$ for $j=1, \cdots, n$, where $C>0$ and $\delta_{1},\delta_{2} \in(0,1)$ are constants independent of $p$, $q$ and $T$.
\end{as1}

The strength of a factor is measured by the number of time series which  are influenced by the factor. As for the common  factors $\mathbf{G}_t$ which is related to most, if
not all, components of $\mathbf{X}_t$, it is reasonable to assume that
\begin{equation}\label{strong}
\begin{split}
\left\|\mathbf{R}_{\cdot i}\right\|^2 \asymp p, \ \ \ \ i=1, \cdots, k_0,\\
\left\|\mathbf{C}_{\cdot j}\right\|^2 \asymp q, \ \ \ \ j=1, \cdots, r_0,
\end{split}
\end{equation}
where $\mathbf{R}_{\cdot i}$ is the $i$-th column of $\mathbf{R}$ and $\mathbf{C}_{\cdot j}$ is the $j$-th column of $\mathbf{C}$, and this is in the same spirit of
the definition for the common factors by \cite{Yao2023}. Let $\mathbf{\Gamma}_{\cdot i}^j$ be the $i$-th column of the $p_j \times k_j$ matrix $\mathbf{\Gamma}_j$ and $\mathbf{\Lambda}_{\cdot i}^j$ be the $i$-th column of the $q_j \times r_j$ matrix $\mathbf{\Lambda}_j$. In the same vein, we further assume that
\begin{equation}\label{weak}
\begin{split}
&\left\|\mathbf{\Gamma}_{\cdot i}^j\right\|^2 \asymp p^{1-\delta_1} \asymp p_j, \ \ \ \ i=1, \cdots, k_j \text { and } j=1, \cdots, m,\\
&\left\|\mathbf{\Lambda}_{\cdot i}^j\right\|^2 \asymp q^{1-\delta_2} \asymp q_j, \ \ \ \ i=1, \cdots, r_j \text { and } j=1, \cdots, n,
\end{split}
\end{equation}
 for the cluster-specific factor loadings in Assumption \ref{ass1}. Note that the factor strength can be measured by the constants $\delta_1,\delta_2 \in(0,1]$ similarly defined in \cite{Lam2012FactorMF}, and (\ref{weak}) indicates that cluster-specific factors are weaker than common factors. 

We use $\text{vec}(\cdot)$ to denote the vectorisation operator that vectoring a matrix by column. Let $\boldsymbol{g}_t:=\text{vec}(\mathbf{G}_t)$,  $\boldsymbol{f}_t:=\text{vec}(\mathbf{F}_t)$  and we denote the 
  lag-(cross)-covariance matrices of the vectorized common factors and cluster-specific factors as follows:
$$
\begin{array}{cl}
\boldsymbol{\Sigma}_g(l)=\text{Cov}\left(\boldsymbol{g}_{t+l}, \boldsymbol{g}_t\right), & \boldsymbol{\Sigma}_f(l)=\text{Cov}\left(\boldsymbol{f}_{t+l}, \boldsymbol{f}_t\right), \\
\boldsymbol{\Sigma}_{g, f}(l)=\text{Cov}\left(\boldsymbol{g}_{t+l}, \boldsymbol{f}_t\right), & \boldsymbol{\Sigma}_{f, g}(l)=\text{Cov}\left(\boldsymbol{f}_{t+l}, \boldsymbol{g}_t\right),
\end{array}
$$
and we assume that all the entries of (cross)-covariance matrices above are bounded and for a pre-determined integer $l_0$,  $\boldsymbol{\Sigma}_g(l)$ and $\boldsymbol{\Sigma}_f(l)$, $l=1,\dots,l_0$ are all full-ranked.
There exist an identifiable issue among the factors and the loading matrices as well recognized in factor models. Let $(\mathbf{U}_1,\mathbf{U}_2)$ be two invertible matrices of sizes $k_0\times k_0$ and $r_0\times r_0$. Then the triplets $(\mathbf{R},\mathbf{G}_t,\mathbf{C})$ and $(\mathbf{RU}_1,\mathbf{U}_1^{-1}\mathbf{G}_t\mathbf{U}_2^{-1},\mathbf{U}_2\mathbf{C})$ are equivalent under model (\ref{model}). Similarly, $(\mathbf{\Gamma},\mathbf{F}_t,\mathbf{\Lambda})$ and $(\mathbf{\Gamma V}_1,\mathbf{V}_1^{-1}\mathbf{F}_t$ $\mathbf{V}_2^{-1},\mathbf{V}_2\mathbf{\Lambda})$ are equivalent as long as $\mathbf{V}_1$ and $\mathbf{V}_2$ are invertible matrices. Model (\ref{model}) is not identifiable, while the linear spaces spanned by the columns of $\mathbf{R}$, $\mathbf{C}$, $\mathbf{\Gamma}$ and $\mathbf{\Lambda}$, denote by $\mathcal{M}(\mathbf{R})$, $\mathcal{M}(\mathbf{C})$, $\mathcal{M}(\mathbf{\Gamma})$ and $\mathcal{M}(\mathbf{\Lambda})$, respectively, can be uniquely determined. To proceed, we decompose the loading matrices as follows,
$$
\begin{aligned}
    \mathbf{R}=\mathbf{P}_1\mathbf{Q}_1,\ \text{and} \ \mathbf{C}=\mathbf{P}_2\mathbf{Q}_2,\\
    \mathbf{\Gamma}=\mathbf{A}_1\mathbf{B}_1,\ \text{and} \ \mathbf{\Lambda}=\mathbf{A}_2\mathbf{B}_2,\\
\end{aligned}
$$
where $\mathbf{P}_i$ and $\mathbf{A}_i$, $i=1,2$ are column orthogonal matrices, $\mathbf{Q}_i$ and $\mathbf{B}_i$, $i=1,2$  are non-singular matrices, i.e.,  $\mathbf{P}_1^{'}\mathbf{P}_1=\mathbf{I}_{k_0}$, $\mathbf{P}_2^{'}\mathbf{P}_2=\mathbf{I}_{r_0}$,
$\mathbf{A}_1^{'}\mathbf{A}_1=\mathbf{I}_{k}$
and $\mathbf{A}_2^{'}\mathbf{A}_2=\mathbf{I}_{r}$. Clearly, $\mathcal{M}(\mathbf{R})=\mathcal{M}(\mathbf{P}_1)$, $\mathcal{M}(\mathbf{C})=\mathcal{M}(\mathbf{P}_2)$, $\mathcal{M}(\mathbf{\Gamma})=\mathcal{M}(\mathbf{A}_1)$ and $\mathcal{M}(\mathbf{\Lambda})=\mathcal{M}(\mathbf{A}_2)$. Let $\mathbf{S}_t=\mathbf{Q}_1\mathbf{G}_t\mathbf{Q}_2^{\top}$ and $\mathbf{V}_t=\mathbf{B}_1\mathbf{F}_t\mathbf{B}_2^{\top}$, then we obtain an alternative formulation of model (\ref{model}) with column-orthonormal loading matrices
\begin{equation*}
\mathbf{X}_t=\mathbf{P}_1\mathbf{S}_t\mathbf{Q}_1^{\top}+\mathbf{A}_1\mathbf{V}_t{\mathbf{A}_2}^{\top}+\mathbf{E}_t^0,\ t=1,\dots,T.
\end{equation*}
 For simplicity, Assumption \ref{ass2} requires that all the loading matrices have orthonormal columns.

\begin{as1}\label{ass2}
 \item[(i)] $\|\mathbf{R}\|_\text{max}=O(p^{-1/2})$, $\|\mathbf{C}\|_\text{max}=O(q^{-1/2})$, 
 $\mathbf{R}^{\top} \mathbf{R}=\mathbf{I}_{k_0}$ and $\mathbf{C}^{\top} \mathbf{C}=\mathbf{I}_{r_0}$; 
 \item[(ii)]$
 \mathbf{\Gamma}_i^{\top} \mathbf{\Gamma}_i=\mathbf{I}_{k_i}$, for $1 \leq i \leq m$, $\mathbf{\Lambda}_j^{\top} \mathbf{\Lambda}_j=\mathbf{I}_{r_j}$, for $1 \leq j \leq n$, $\|{\boldsymbol{\gamma}}_i\|^2\asymp p^{\delta_1-1}$ for $1\leq i\leq p$, $\|{\boldsymbol{\lambda}}_j\|^2\asymp q^{\delta_2-1}$ for $1\leq i\leq q$, where ${\boldsymbol{\gamma}}^{\top}_i$ and ${\boldsymbol{\lambda}}^{\top}_j$ denote the i-th row of matrix $\mathbf{\Gamma}$ and j-th row of matrix $\mathbf{\Lambda}$, respectively;
 \item[(iii)] assume that there exists
 a constant $c_0 \in(0,1)$ such that
\begin{equation}\label{26}
\begin{split}
\left\|\mathbf{RR}^{\top}\mathbf{\Gamma}\right\| \leq c_0,\
\left\|\mathbf{CC}^{\top}\mathbf{\Lambda}\right\| \leq c_0.
\end{split}
\end{equation}
\item[(iv)]
assume that  neither $ \operatorname{rp}\left(\mathbf{\Gamma}_i\right)\left(\operatorname{rp}\left(\mathbf{\Gamma}_i\right)\right)^{\top}$ nor $ \operatorname{rp}\left(\mathbf{\Lambda}_j\right)\left(\operatorname{rp}\left(\mathbf{\Lambda}_j\right)\right)^{\top}$ for $i=1, \cdots, m$, $ j=1, \cdots, n$ can be written  as a block diagonal matrix with at least two blocks, where $\operatorname{rp}\left(\mathbf{\Gamma}_i\right)$ and $\operatorname{rp}\left(\mathbf{\Lambda}_j\right)$ denote any row-permutation of $\mathbf{\Gamma}_i$ and $\mathbf{\Lambda}_j$, respectively.
\end{as1}


Although the loading matrices $\mathbf{R}$, $\mathbf{C}$, $\mathbf{\Gamma}$ and $\mathbf{\Lambda}$ are still not unique under Assumption \ref{ass2}, the factor loading spaces are uniquely identifiable by Model (\ref{model}). Assumption \ref{ass2}.($iii$) implies that the columns of row/column local loading matrices do not fall entirely into the spaces spanned by the columns of global row/column loading matrices, respectively. Assumption  \ref{ass2}.($iv$) ensures that the numbers of clusters $m$ and $n$ are uniquely defined.

We also assume that the noises are independent of the factors and allow the dependence between the common factors and cluster-specific factors; see Assumption \ref{ass3} below. In contrast to the cases where the latent factors capture all the dynamic dependence of $\mathbf{X}_t$'s, i.e., there exist no serial dependence in  the  noise process,  weak serial dependence is allowed for the idiosyncratic noise process in our paper, and the factors can be dynamically dependent; see Assumption \ref{ass5} in Section \ref{sec3}. 

\begin{as1}\label{ass3}
	Let $\{\mathbf{X}_t\}, \{\mathbf{G}_t\}$ and $\{\mathbf{F}_t\}$ be strictly stationary with  finite fourth moments. As $p,q \rightarrow \infty$, it holds that for $l=0,1, \cdots, l_0$,
	$$
	\begin{gathered}
		\left\|\boldsymbol{\Sigma}_g(l)\right\| \asymp pq \asymp\left\|\boldsymbol{\Sigma}_g(l)\right\|_{\min }, \\
		\left\|\boldsymbol{\Sigma}_f(l)\right\| \asymp p^{1-\delta_1}q^{1-\delta_2} \asymp\left\|\boldsymbol{\Sigma}_f(l)\right\|_{\min }, \\
		\left\|\boldsymbol{\Sigma}_g(l)^{-1 / 2} \boldsymbol{\Sigma}_{g, f}(l) \boldsymbol{\Sigma}_f(l)^{-1 / 2}\right\| \leq c_0<1, \quad\left\|\boldsymbol{\Sigma}_f(l)^{-1 / 2} \boldsymbol{\Sigma}_{f, g}(l) \boldsymbol{\Sigma}_g(l)^{-1 / 2}\right\| \leq c_0<1, \\
		\left\|\boldsymbol{\Sigma}_{g, f}(l)\right\|=O\left(p^{1-\delta_1 / 2}q^{1-\delta_2 / 2}\right), \quad\left\|\boldsymbol{\Sigma}_{f, g}(l)\right\|=O\left(p^{1-\delta_1 / 2}q^{1-\delta_2 / 2}\right).
	\end{gathered}
	$$
	Furthermore, assume that ${\text{Cov}}\left(\boldsymbol{g}_{t}, \boldsymbol{e}_s^0\right)=\boldsymbol{0}, {\text{Cov}}\left(\boldsymbol{f}_{t}, \boldsymbol{e}_s^0\right)=\boldsymbol{0}$ for any $t$ and $s$, where $\boldsymbol{e}_s^0=vec(\mathbf{E}_s^0)$.
\end{as1}

In this study, our goal is to recover the latent row/column clusters simultaneously for matrix-valued time series under Model (\ref{model}) with both strong/global and weak/cluster-specific factors, and the detailed procedures will be elaborated on in the following subsections. A naive way is to vectorize the matrices into long vectors and then cluster a large number of time series into different and unknown clusters such that the members within each cluster share a similar dynamic structure, but this would ignore the matrix structures and lead to both the increase of computational burden and statistical efficiency loss, which has been well recognized in many statistical learning problems for matrix-valued data \citep{liu2023simultaneous,Fan2023,he2024online}. In addition, for biclustering problems, the naive vectorization method would be a disaster as the row/column variables are mixed together and there is no guarantee that any vector clustering method would achieve separable clustering results for row/column variables.
The new matrix factor model in (\ref{model}) not only reveals the serial dynamics for a panel of row/column variables, but also explores the spatial correlations among entries of the observation matrix in a parsimonious way, which provide a flexible way to separate those latent row  and column clusters from observations \(\{\mathbf{X}_t\}_{1\le t\le T}\). 

In the following, we elaborate on the detailed biclustering procedures based on Model (\ref{model}). Firstly, the  numbers of global and cluster-specific factors are  unknown and need to be estimated in advance, which will be discussed in Section \ref{sec2.1}. Then we discuss the estimation of the  loading  matrices of global/common factors in Section \ref{sec2.2} and that of cluster-specific factors in Section \ref{sec2.3}. At last, we provide the $K$-means algorithm based on the estimated cluster-specific loading matrices in Section \ref{sec2.4}.
\subsection{Estimation for the number of factors}\label{sec2.1}
 It’s well known that accurate estimation of the numbers of factors is of great importance to do matrix factor analysis. Firstly, we estimate the numbers of row and column common factors by an one-pass, eigenvalue ratio-based method. 
Let $l$ be a positive integer and the $j$-th column of $\mathbf{X}_t$ be $\mathbf{x}_{t,.j}$ for \(j=1,\dots,q\). Define
\begin{equation}
    \mathbf{\Sigma}_{x,ij}(l)=\text{Cov}(\mathbf{x}_{t,.i},\mathbf{x}_{t+l,.j}), \ \ \text{for} \ \ i,j=1,\dots,q.
\end{equation}
For a pre-determined integer \(l_0\ge 0\), define
\begin{equation}
    \mathbf{M}_{0,1}=\sum_{l=1}^{l_0}\sum_{i=1}^{q}\sum_{j=1}^{q}\mathbf{\Sigma}_{x,ij}(l)\mathbf{\Sigma}_{x,ij}^{\top}(l).
\end{equation}
Recall the zero mean assumptions of factors and idiosyncratic component, we define the sample version of $\mathbf{\Sigma}_{x,ij}(l)$ and $\mathbf{M}_{0,1}$ as follows
\begin{equation}\label{sigmahatx}
\widehat{\mathbf{\Sigma}}_{x,ij}(l)=\frac{1}{T-l}\sum_{t=1}^{T-l}\mathbf{x}_{t,.i}\mathbf{x}^{\top}_{t+l,.j},
\end{equation}
\begin{equation}\label{sigmahatM01}
\widehat{\mathbf{M}}_{0,1}=\sum_{l=1}^{l_0}\sum_{i=1}^{q}\sum_{j=1}^{q}\widehat{\mathbf{\Sigma}}_{x,ij}(l)\widehat{\mathbf{\Sigma}}_{x,ij}^{\top}(l).
\end{equation}

Let $\widehat{\lambda}_{1,1} \geq \widehat{\lambda}_{1,2} \geq \cdots \geq \widehat{\lambda}_{1, p} \geq 0$ be the ordered eigenvalues of $\widehat{\mathbf{M}}_{0,1}$. For a pre-specified positive integer \(J_0\le p\), we define
\begin{eqnarray}\label{29}
	&\widehat{R}_{1,j}=\widehat{\lambda}_{1,j}/\widehat{\lambda}_{1,j+1},\ 1\le j\le J_0-1,
\end{eqnarray}
then we let \(\widehat{R}_{\widehat{\mu}_1}\) and \(\widehat{R}_{\widehat{\mu}_2}\) be  the two largest local maximum among \(\widehat{R}_{1,1},\dots,\widehat{R}_{1,J_0-1}\). The estimators for the numbers of row factors are then defined as
\begin{eqnarray}\label{210}
	&\widehat{k}_0=\text{min}\{\widehat{\mu}_1,\widehat{\mu}_2\},\ \widehat{k}_0+\widehat{k}=\text{max}\{\widehat{\mu}_1,\widehat{\mu}_2\}.
\end{eqnarray}

We use the ratios of the cumulative eigenvalues in (\ref{29}) to add together the information from different lags. As the autocorrelation is often at its strongest at small time lags, we usually take $l$ as a smaller integer such as $l=1,\dots,5$ in simulations and practical applications.

Many approaches to identify the number of factors have been developed in the literature. Factor model is in essence characterized by the presence of a large eigengap between eigenvalues of the covariance matrix \citep{barigozzi2020consistent}. For instance, \cite{Wang2019,Fan2023,He2024Matrix} use the ratios of the ordered eigenvalues of $\widehat{\mathbf{M}}_{0,1}$ and show that the estimated eigenvalue ratio will drop sharply at a certain point. However, the aforementioned methods cannot estimate the numbers of strong and all the weaker factors for Model (\ref{model}) in a manner of one-pass. We handle this situation with a similar idea in \cite{Yao2023} and obtain consistent ratio-based estimators; see Theorem \ref{facnumthm} in Section \ref{sec3}.

For $\widehat{r}_0$ and $\widehat{r}$, they can be estimated by performing the same procedure on the transposes of $\mathbf{X}_t$'s to construct $\mathbf{M}_{0,2}$ and $\widehat{\mathbf{M}}_{0,2}$ as follows:
\begin{equation}\label{M02} \mathbf{M}_{0,2}=\sum_{l=1}^{l_0}\sum_{i=1}^{q}\sum_{j=1}^{q}\mathbf{\Sigma}_{x',ij}(l)\mathbf{\Sigma}_{x',ij}^{\top}(l),
\end{equation}
\begin{equation}\label{M02hat}
\widehat{\mathbf{M}}_{0,2}=\sum_{l=1}^{l_0}\sum_{i=1}^{q}\sum_{j=1}^{q}\widehat{\mathbf{\Sigma}}_{x',ij}(l)\widehat{\mathbf{\Sigma}}_{x',ij}^{\top}(l).
\end{equation}
where $\mathbf{\Sigma}_{x',ij}(l)=\text{Cov}(\mathbf{X}_{t,i.},\mathbf{X}_{t+l,j.})$ and $\widehat{\mathbf{\Sigma}}_{x',ij}(l)=\frac{1}{T-l}\sum_{t=1}^{T-l}\mathbf{X}_{t,i.}\mathbf{X}^{\top}_{t+l,j.}$. By utilizing the eigenvalue ratio of $\widehat{\mathbf{M}}_{0,2}$ to determine the local maximum $\widehat{R}_{2,j}=\widehat{\lambda}_{2,j}/\widehat{\lambda}_{2,j+1}$, let \(\widehat{R}_{\widehat{\kappa}_1}\) and \(\widehat{R}_{\widehat{\kappa}_2}\) be the the two largest local maximum among \(\widehat{R}_{2,1},\dots,\widehat{R}_{2,J_0-1}\). The estimators for the numbers of strong and weak column factors are defined as
\begin{eqnarray}\label{facnum_r}
	&\widehat{r}_0=\text{min}\{\widehat{\kappa}_1,\widehat{\kappa}_2\},\ \widehat{r}_0+\widehat{r}=\text{max}\{\widehat{\kappa}_1,\widehat{\kappa}_2\}.
\end{eqnarray}
\begin{rmk1}
\begin{enumerate}
    \item[(i)] Note that estimators in (\ref{210}) and (\ref{facnum_r}) are derived under the assumption that all the cluster-specific factors in one direction are of the same factor strength.
    \item[(ii)] In order to mitigate the impact of the indeterminate form "0/0", we truncate the sequence at index $J_0$. In practice, we choose an appropriate finite value for $J_0$ based on the size of $p$ and $q$.
\end{enumerate}
\end{rmk1}

\subsection{Estimation for the loading matrices of common factors}\label{sec2.2}
In this section, we propose a projection  method to estimate the factor loading matrices corresponding to the common factors $\mathbf{G}_t$. In the following we assume that the number of common and cluster-specific factors are given in advance, otherwise we can estimate them using the technique introduced in Section \ref{sec2.1}.  
We first absorb the cluster-specific term into the idiosyncratic error term and then propose a similar projection technique as \cite{YU2022} to estimate the global loading matrices, except that our estimation is based on  auto-cross-covariances in contrast to covariance matrices in \cite{YU2022}.
In detail, rewrite Model (\ref{model}) as	\begin{equation}\label{main2}
        \mathbf{X}_t=\mathbf{R}\mathbf{G}_t\mathbf{C}^{\top}+{\mathbf{E}}_t,
		\end{equation}
		where \({\mathbf{E}}_t=\mathbf{\Gamma}\mathbf{F}_t{\mathbf{\Lambda}}^{\top}+\mathbf{E}_t^0\), then the observation matrices \(\mathbf{X}_t\) satisfy a standard matrix factor model  (\ref{factor_model}) in the literature \citep{Wang2019,Fan2023,He2023Vector}.
		We first assume that $\mathbf{R}$ and $\mathbf{C}$ are known, and later give initial estimates of  $\mathbf{R}$ and $\mathbf{C}$.
        Let $\mathbf{Z}_t=\mathbf{X}_t\mathbf{C}$ and $\mathbf{W}_t=\mathbf{X}_t^{\top}\mathbf{R}$, that is, we project the observations onto the space spanned by the columns of $\mathbf{C}$ and $\mathbf{R}$, i.e., $\mathcal{M}(\mathbf{C})$ and $\mathcal{M}(\mathbf{R})$ first. Note that $\mathbf{R}$ and $\mathbf{C}$ are column-orthogonal matrices. It follows from model (\ref{main2}) that
$$
\begin{aligned}
\mathbf{Z}_t=&\mathbf{X}_t\mathbf{C}=\mathbf{R}\mathbf{G}_t+{\mathbf{E}}_t\mathbf{C},\\
\mathbf{W}_t=&\mathbf{X}_t^{\top}\mathbf{R}=\mathbf{C}\mathbf{G}_t^{\top}+{\mathbf{E}}_t^{\top}\mathbf{R},
\end{aligned}
$$
where $\Zb_t$'s and $\Wb_t$'s are respectively $p\times r_0$ and $q\times k_0$ matrix-valued observations, with column dimension much lower than that of $\Xb_t$ and $\Xb_t^{\top}$. 

Let the $j$-th column of $\mathbf{Z}_t$ ($\mathbf{W}_t$) be $\mathbf{z}_{t,.j}$ ($\mathbf{w}_{t,.j}$). The columns of $\mathbf{Z}_t$ and $\mathbf{W}_t$ can be written in the form of a vector factor model as:
  $$
  \begin{aligned}
      \mathbf{z}_{t,.j}&=\mathbf{R}\mathbf{G}_{t,.j}+{\mathbf{E}}_t\mathbf{C}_{.j}:=\mathbf{R}\bar{\boldsymbol{g}}_{t,j}+\bar{\boldsymbol{e}}_{t,j},\\
      \mathbf{w}_{t,.i}&=\mathbf{C}\mathbf{G}_{t,i.}+{\mathbf{E}}_t^{\top}\mathbf{R}_{.i}:=\mathbf{C}\dot{\boldsymbol{g}}_{t,i}+\dot{\boldsymbol{e}}_{t,i}.
  \end{aligned}
  $$
  Define $\mathbf{\Sigma}_{z,ij}(l)=\text{Cov}(\mathbf{z}_{t,.i},\mathbf{z}_{t+l,.j})$ and $\mathbf{\Sigma}_{\bar{g},ij}(l)=\text{Cov}(\boldsymbol{\bar{g}}_{t,.i},\boldsymbol{\bar{g}}_{t+l,.j})$, then we can construct $\mathbf{M}_{1}$ as follows:
\begin{equation}\label{M1}
\mathbf{M}_{1}=\sum_{l=1}^{l_0}\sum_{i=1}^{r_0}\sum_{j=1}^{r_0}\mathbf{\Sigma}_{z,ij}(l)\mathbf{\Sigma}_{z,ij}^{\top}(l)\approx\mathbf{R}\left(\sum_{l,i,j}\mathbf{\Sigma}_{\bar{g},ij}(l)\mathbf{\Sigma}_{\bar{g},ij}^{\top}(l)\right)\mathbf{R}^\top,
\end{equation}
where \(l_0\) is a pre-determined integer and $\mathbf{R}\left(\sum_{l,i,j}\mathbf{\Sigma}_{\bar{g},ij}(l)\mathbf{\Sigma}_{\bar{g},ij}^{\top}(l)\right)\mathbf{R}^\top$ is the leading term of $\mathbf{M}_1$ based on Assumptions \ref{ass2} and \ref{ass3}.

By a similar argument, we can construct $\mathbf{\Sigma}_{w,ij}(l)=\text{Cov}(\mathbf{w}_{t,.i},\mathbf{w}_{t+l,.j})$ and $\mathbf{M}_{2}$ as follows:
\begin{equation}\label{M2}
\mathbf{M}_{2}=\sum_{l=1}^{l_0}\sum_{i=1}^{k_0}\sum_{j=1}^{k_0}\mathbf{\Sigma}_{w,ij}(l)\mathbf{\Sigma}_{w,ij}^{\top}(l)
\end{equation}
Assume that $\mathbf{M}_1$ ($\mathbf{M}_2$) has at least $k_0$ ($r_0$) distinct nonzero eigenvalues. Then, the factor loading space $\mathcal{M}(\mathbf{R})$ ($\mathcal{M}(\mathbf{C})$) can be estimated by the space spanned by the eigenvectors corresponding to the $k_0$ ($r_0$) largest non-zero eigenvalues of $\mathbf{M}_1$ ($\mathbf{M}_2$). 

One problem of the above ideal argument is that the projection matrix $\mathbf{R}$ and $\mathbf{C}$ are not available in practice. A natural solution is to replace them with their consistent initial estimates. Therefore, we use the leading \(k_0\) eigenvectors of $\widehat{\mathbf{M}}_{0,1}$, which are defined in (\ref{sigmahatM01}), as an initial estimator of $\mathbf{R}$, denoted as \(\widehat{\mathbf{R}}^{0}\). Similarly, we construct \(\widehat{\mathbf{C}}^{0}\) by the leading \(r_0\) eigenvectors of $\widehat{\mathbf{M}}_{0,2}$ defined in (\ref{M02hat}).

Finally, we construct  $\widehat{\mathbf{M}}_{1}$ based on the projected data 
$\widehat{\mathbf{Z}}_t:=\mathbf{X}_t\widehat{\mathbf{C}}^{0}$.
 Define $$\widehat{\boldsymbol{\Sigma}}_{\widehat{z},ij}(l)=\frac{1}{T-l}\sum_{t=1}^{T-l}\widehat{\mathbf{Z}}_{t,.i}\widehat{\mathbf{Z}}^{\top}_{t+l,.j}$$ and further construct $\widehat{\mathbf{M}}_{1}$ as
  \begin{equation}\label{M1hat}
\widehat{\mathbf{M}}_{1}=\sum_{l=1}^{l_0}\sum_{i=1}^{r_0}\sum_{j=1}^{r_0}\widehat{\boldsymbol{\Sigma}}_{\widehat{z},ij}(l)\widehat{\boldsymbol{\Sigma}}_{\widehat{z},ij}(l)^{\top},
  \end{equation}
then we obtain the estimator of global row loading  matrix $\widehat{\mathbf{R}}$,  whose columns are the top $k_0$ eigenvectors of $\widehat{\mathbf{M}}_{1}$. At the same time,  define $\widehat{\mathbf{M}}_2$ in a similar way with transposes of $\mathbf{X}_t$'s, i.e., let
\begin{equation}\label{M2hat}
\widehat{\mathbf{M}}_{2}=\sum_{l=1}^{l_0}\sum_{i=1}^{k_0}\sum_{j=1}^{k_0}\widehat{\mathbf{\Sigma}}_{\widehat{w},ij}(l)\widehat{\mathbf{\Sigma}}_{\widehat{w},ij}^{\top}(l),
\end{equation}
where $\widehat{\mathbf{W}}_t=\mathbf{X}_t^{\top}\widehat{\mathbf{R}}^{0}$ and $\widehat{\boldsymbol{\Sigma}}_{\widehat{w},ij}(l)=\frac{1}{T-l}\sum_{t=1}^{T-l}\widehat{\mathbf{W}}_{t,.i}\widehat{\mathbf{W}}^{\top}_{t+l,.j}$. Then the global column loading matrix $\mathbf{C}$ is estimated by the leading $r_0$ eigenvectors of $\widehat{\mathbf{M}}_2$. For better illustration, we summarize the procedure for estimating the global matrix factor spaces in Algorithm \ref{alg1}.
\begin{algorithm}[h]
\caption{{Estimation procedure for the global loading matrices} \label{alg1}}
\KwIn{Data matrices $\{\mathbf{X}_t\}_{t\leq T}$, the strong factor numbers of row and column $(k_0,r_0)$, a positive integer $l_0$.}
\KwOut{Estimation for the global loading matrices $\widehat{\mathbf{R}}$ and $\widehat{\mathbf{C}}$.}
\begin{algorithmic}[1]
\State given data matrix $\mathbf{X}_t$, for $l=1,\dots,l_0$, define $\widehat{\mathbf{M}}_{0,1}$ and $\widehat{\mathbf{M}}_{0,2}$ in (\ref{sigmahatM01}) and (\ref{M02hat}), obtain the initial estimators of global loading matrices by the leading $k_0$ and $r_0$ eigenvectors of $\widehat{\mathbf{M}}_{0,1}$ and $\widehat{\mathbf{M}}_{0,2}$ respectively, denoted as $\widehat{\mathbf{R}}^0$ and $\widehat{\mathbf{C}}^0$;
\State project the observations to  lower dimensions by letting $\widehat{\mathbf{Z}}_t:=\mathbf{X}_t\widehat{\mathbf{C}}^{0}$, $\widehat{\mathbf{W}}_t=\mathbf{X}_t^{\top}\widehat{\mathbf{R}}^{0}$, then define $\widehat{\mathbf{M}}_{1}$ and $\widehat{\mathbf{M}}_{2}$ as (\ref{M1hat}) and (\ref{M2hat});
\State the estimators of global row and column loading matrices  $\widehat{\mathbf{R}}$ and $\widehat{\mathbf{C}}$ are finally given by the leading $k_0$ and $r_0$ eigenvectors of $\widehat{\mathbf{M}}_1$ and $\widehat{\mathbf{M}}_2$, respectively.
\end{algorithmic}
\end{algorithm}
\subsection{Estimation for the loading matrices of cluster-specific factors}\label{sec2.3}
In this section, we focus on estimating the loading matrices corresponding to the cluster-specific factors. As \cite{Lam2012FactorMF} shows that weak factors in vector factor model can be more accurately estimated by removing the effect of strong factors from the data, we first remove the effect of the common factors part $\mathbf{R}\mathbf{G}_t\mathbf{C}^{\top}$ from the observations before estimating $\mathbf{\Gamma}$ and $\mathbf{\Lambda}$. To remove the effect of the common factors, we propose to project the observation matrices onto the orthogonal complement space of the global loading spaces. In detail, we define \begin{equation}\label{equ:Y}
    \mathbf{Y}_t=(\mathbf{I}_p-{\mathbf{R}}{\mathbf{R}}^{\top})\mathbf{X}_t(\mathbf{I}_q-{\mathbf{C}}{\mathbf{C}}^{\top})={\mathbf{P}}_{{R}}^c\mathbf{X}_t{\mathbf{P}}_{{C}}^c.
\end{equation}
 After transformation, \(\mathbf{Y}_t\) follows a standard matrix factor model (\ref{factor_model}), and \(\mathbf{F}_t\) can be regarded as common matrix factors, i.e., 
 \begin{equation}
     \mathbf{Y}_t={\mathbf{P}}_{{R}}^c\mathbf{\Gamma}\mathbf{F}_t\mathbf{\Lambda}^{\top}{\mathbf{P}}_{{C}}^c+{\mathbf{P}}_{{R}}^c\mathbf{E}^0_t{\mathbf{P}}_{{C}}^c.
 \end{equation}
 Given $\mathbf{Y}_t$, we can adopt a similar projection estimation method as introduced in Section \ref{sec2.2} to estimate the matrices corresponding to the cluster-
specific factors.
 Specifically, we set $\mathbf{U}_t=\mathbf{Y}_t\mathbf{\Lambda}$, $ \mathbf{H}_t=\mathbf{Y}_t^{\top}\mathbf{\Gamma}$,
 where $\mathbf{U}_t$ is a $p\times r$ matrix-valued observation and $\mathbf{H}_t$ is a $q\times k$ matrix-valued observation. For the $i$-th column of $\mathbf{U}_t$ and $\mathbf{H}_t$, i.e., $\mathbf{u}_{t,.i}$ and $\mathbf{h}_{t,.i}$, define
 \begin{equation*}
     \mathbf{\Sigma}_{u,ij}(l)=\text{Cov}(\mathbf{u}_{t,.i},\mathbf{u}_{t+l,.j}), \quad i,j=1,\dots,r.
 \end{equation*}
  \begin{equation*}
     \mathbf{\Sigma}_{h,ij}(l)=\text{Cov}(\mathbf{h}_{t,.i},\mathbf{h}_{t+l,.j}), \quad i,j=1,\dots,k.
 \end{equation*}
		Similar to the definition of ${\mathbf{M}}_1$ in (\ref{M1}), we define ${\mathbf{M}}^{*}_1$ as
\begin{equation*}
    {\mathbf{M}}^{*}_1=\sum_{l=1}^{l_0}\sum_{i=1}^{r}\sum_{j=1}^{r}\mathbf{\Sigma}_{u,ij}(l)\mathbf{\Sigma}_{u,ij}^{\top}(l).
\end{equation*}
  We estimate the cluster-specific row loading matrix $\mathbf{\Gamma}$ by the eigenvectors corresponding to the none-zero eigenvalues of ${\mathbf{M}}^{*}_1$. Similarly, to estimate the cluster-specific column loading matrix $\mathbf{\Lambda}$, we first construct
\begin{equation*}
    {\mathbf{M}}^{*}_2=\sum_{l=1}^{l_0}\sum_{i=1}^{k}\sum_{j=1}^{k}\mathbf{\Sigma}_{h,ij}(l)\mathbf{\Sigma}_{h,ij}^{\top}(l)
\end{equation*}
and then calculate its eigenvectors corresponding to the leading $k$ eigenvalues.

The projection matrices \({\mathbf{\Gamma}}\) and \({\mathbf{\Lambda}}\) are also not available, we replace $\mathbf{R}$ ($\mathbf{C}$) with $\hat{\mathbf{R}}$ ($\hat{\mathbf{C}}$) defined in Section \ref{sec2.2} to get initial estimators $\hat{\mathbf{\Gamma}}^0$ and $\hat{\mathbf{\Lambda}}^0$. In detail, let
  \begin{equation*}
      \widehat{\mathbf{Y}}_t=(\mathbf{I}_p-\widehat{\mathbf{R}}\widehat{\mathbf{R}}^{\top})\mathbf{X}_t(\mathbf{I}_p-\widehat{\mathbf{C}}\widehat{\mathbf{C}}^{\top})={\mathbf{P}}_{\widehat{R}}^c\mathbf{X}_t{\mathbf{P}}_{\widehat{C}}^c,
  \end{equation*}
and denote the $j$-th column of $\widehat{\mathbf{Y}}_t$ as $\widehat{\mathbf{y}}_{t,.j}$ and construct $\widehat{\mathbf{M}}^{*}_{0,1}$ as follows
\begin{equation}\label{M01star}
    \widehat{\mathbf{M}}^{*}_{0,1}=\sum_{l=1}^{l_0}\sum_{i=1}^{q}\sum_{j=1}^{q}\widehat{\mathbf{\Sigma}}_{\hat{y},ij}(l)\widehat{\mathbf{\Sigma}}_{\hat{y},ij}^{\top}(l),
    \end{equation}
where $\widehat{\mathbf{\Sigma}}_{\hat{y},ij}(l)=\frac{1}{T-l}\sum_{t=1}^{T-l}\mathbf{\hat{y}}_{t,.i}\mathbf{\hat{y}}^{\top}_{t+l,.j}$. We prove that the leading $k$ eigenvectors of $\widehat{\mathbf{M}}^{*}_{0,1}$ lie in the same column space of $(\mathbf{I}_p-{\mathbf{R}}{\mathbf{R}}^{\top})\mathbf{\Gamma}$ asymptotically under mild conditions. Therefore, we use the leading $k$ eigenvectors of $\widehat{\mathbf{M}}^{*}_{0,1}$ as an initial estimator of $(\mathbf{I}_p-{\mathbf{R}}{\mathbf{R}}^{\top})\mathbf{\Gamma}$, denoted as $\widehat{\mathbf{\Gamma}}^{0}$. 

Similarly we  apply the same procedure to the transposes of $\{\widehat{\mathbf{Y}}_t,t=1,\dots,T \}$ and construct $\widehat{\mathbf{\Sigma}}_{\hat{y}',ij}(l)=\frac{1}{T-l}\sum_{t=1}^{T-l}\mathbf{\hat{Y}}_{t,i.}\mathbf{\hat{Y}}^{\top}_{t+l,j.}$ and $\widehat{\mathbf{M}}^{*}_{0,2}$ as follows,
 \begin{equation}\label{M02starhat}
    \widehat{\mathbf{M}}^{*}_{0,2}=\sum_{l=1}^{l_0}\sum_{i=1}^{p}\sum_{j=1}^{p}\widehat{\mathbf{\Sigma}}_{\hat{y}',ij}(l)\widehat{\mathbf{\Sigma}}_{\hat{y}',ij}^{\top}(l).
\end{equation}
We estimate the local column loading matrix $(\mathbf{I}_q-{\mathbf{C}}{\mathbf{C}}^{\top})\mathbf{\Lambda}$ by $\widehat{\mathbf{\Lambda}}^{0}$ whose columns are the normalized eigenvectors corresponding to the $r$ largest eigenvalues of $\widehat{\mathbf{M}}^{*}_{0,2}$. Based on the projected data $\widehat{\mathbf{U}}_t:=\widehat{\mathbf{Y}}_t\widehat{\mathbf{\Lambda}}^{0}$ ($\widehat{\mathbf{H}}_t:=\widehat{\mathbf{Y}}^\top_t\widehat{\mathbf{\Gamma}}^{0}$), we further construct the sample versions of $\mathbf{M}^*_1$ ($\mathbf{M}^*_2$), denoted as $\widehat{\mathbf{M}}^*_1$ ($\widehat{\mathbf{M}}^*_2$), where
\begin{equation*}\label{M1starhat}
    \widehat{\mathbf{\Sigma}}_{\hat{u},ij}(l)=\frac{1}{T-l}\sum_{t=1}^{T-l}\left(\widehat{\mathbf{Y}}_t\widehat{\mathbf{\Lambda}}^{0}_{.i}\widehat{\mathbf{\Lambda}}^{0\top}_{.j}\widehat{\mathbf{Y}}_{t+l}^{\top}\right),\ \ \widehat{\mathbf{M}}^{*}_{1}=\sum_{l=1}^{l_0}\sum_{i=1}^{q}\sum_{j=1}^{q}\widehat{\mathbf{\Sigma}}_{\hat{u},ij}(l)\widehat{\mathbf{\Sigma}}_{\hat{u},ij}^{\top}(l).
\end{equation*}
\begin{equation*}\label{M2starhat}
    \widehat{\mathbf{\Sigma}}_{\hat{h},ij}(l)=\frac{1}{T-l}\sum_{t=1}^{T-l}\left(\widehat{\mathbf{Y}}_t^{\top}\widehat{\mathbf{\Gamma}}^{0}_{.i}\widehat{\mathbf{\Gamma}}^{0\top}_{.j}\widehat{\mathbf{Y}}_{t+l}\right),\ \ \widehat{\mathbf{M}}^{*}_{2}=\sum_{l=1}^{l_0}\sum_{i=1}^{p}\sum_{j=1}^{p}\widehat{\mathbf{\Sigma}}_{\hat{h},ij}(l)\widehat{\mathbf{\Sigma}}_{\hat{h},ij}^{\top}(l).
\end{equation*}
We obtain the top $k$ ($r$) eigenvalues of $\widehat{\mathbf{M}}^{*}_{1}$ ($\widehat{\mathbf{M}}^{*}_{2}$) and denote the corresponding eigenvectors by $\widehat{\boldsymbol{\zeta}}_{1,1},\dots,\widehat{\boldsymbol{\zeta}}_{1,k}$ ($\widehat{\boldsymbol{\zeta}}_{2,1},\dots,\widehat{\boldsymbol{\zeta}}_{2,r}$). The estimators of the local loading matrices for the cluster-specific factors are constructed as 
\begin{equation*}
    \begin{aligned}
      \widehat{\mathbf{\Gamma}}=(\widehat{\boldsymbol{\zeta}}_{1,1},\dots,\widehat{\boldsymbol{\zeta}}_{1,k}), 
      \ \ \ \widehat{\mathbf{\Lambda}}=(\widehat{\boldsymbol{\zeta}}_{2,1},\dots,\widehat{\boldsymbol{\zeta}}_{2,r}).
    \end{aligned}
\end{equation*}
In fact,  $\mathcal{M}(\widehat{\mathbf{R}})$ and $\mathcal{M}(\widehat{\mathbf{C}})$ are  consistent estimators for $\mathcal{M}({\mathbf{R}})$ and $\mathcal{M}({\mathbf{C}})$ under mild conditions, respectively. However $\mathcal{M}(\widehat{\mathbf{\Gamma}})$ and $\mathcal{M}(\widehat{\mathbf{\Lambda}})$ are consistent estimators for $\mathcal{M}\left((\mathbf{I}_p-{\mathbf{R}}{\mathbf{R}}^{\top})\mathbf{\Gamma}\right)$ and $\mathcal{M}\left((\mathbf{I}_q-{\mathbf{C}}{\mathbf{C}}^{\top})\mathbf{\Lambda}\right)$, respectively, rather than  estimators of $\mathcal{M}({\mathbf{\Gamma}})$ or $\mathcal{M}({\mathbf{\Lambda}})$, see Theorem \ref{th2} in Section \ref{sec3} for detailed consistency results. However, this would not bring any trouble to the biclustering task for matrix-valued time series. 
For notation simplicity, we denote $(\mathbf{I}_p-{\mathbf{R}}{\mathbf{R}}^{\top})\mathbf{\Gamma}=\mathbf{P}_{\mathbf{R}\bot\mathbf{\Gamma}}$ and $(\mathbf{I}_p-{\mathbf{C}}{\mathbf{C}}^{\top})\mathbf{\Lambda}=\mathbf{P}_{\mathbf{C}\bot\mathbf{\Lambda}}$.

Finally, we summarize the procedures to estimate the loadings for cluster-specific factors in the following  Algorithm \ref{alg2}.
\begin{algorithm}[h]
\caption{{Estimation for the cluster-specific loading matrices} \label{alg2}}
\KwIn{Data matrices $\{\mathbf{X}_t\}_{t\leq T}$, the strong factor numbers of row and column $(k_0,r_0)$, the weak factor numbers of row and column $(k,r)$, a positive integer $l_0$.}
\KwOut{Estimators of the cluster-specific loading matrices $\widehat{\mathbf{\Gamma}}$ and $\widehat{\mathbf{\Lambda}}$.}
\begin{algorithmic}[1]
\State given data matrix $\mathbf{X}_t$, for $l=1,\dots,l_0$, obtain the final estimators of global loading matrices $\widehat{\mathbf{R}}$ and $\widehat{\mathbf{C}}$ by Algorithm \ref{alg1};
\State project the data matrices to the orthogonal complement space of the global loading spaces by defining $\widehat{\mathbf{Y}}_t={\mathbf{P}}_{\widehat{R}}^c\mathbf{X}_t{\mathbf{P}}_{\widehat{C}}^c$;
\State given $\widehat{\mathbf{Y}}_t$, define
$\widehat{\mathbf{M}}^{*}_{0,1}$ and 
$\widehat{\mathbf{M}}^{*}_{0,2}$ by equations (\ref{M01star}) and (\ref{M02starhat}), use the leading $k$ ($r$) eigenvectors of $\widehat{\mathbf{M}}^{*}_{0,1}$ ($\widehat{\mathbf{M}}^{*}_{0,2}$) as an initial estimator of $(\mathbf{I}_p-{\mathbf{R}}{\mathbf{R}}^{\top})\mathbf{\Gamma}$ ($(\mathbf{I}_q-{\mathbf{C}}{\mathbf{C}}^{\top})\mathbf{\Lambda}$), denoted as $\widehat{\mathbf{\Gamma}}^{0}$ ($\widehat{\mathbf{\Lambda}}^{0}$); 
\State construct the projected data $\widehat{\mathbf{U}}_t:=\widehat{\mathbf{Y}}_t\widehat{\mathbf{\Lambda}}^{0}$ and $\widehat{\mathbf{H}}_t:=\widehat{\mathbf{Y}}^\top_t\widehat{\mathbf{\Gamma}}^{0}$, further calculate
$\widehat{\mathbf{M}}^{*}_{1}$ and 
$\widehat{\mathbf{M}}^{*}_{2}$ using (\ref{M1starhat}) and (\ref{M2starhat});
\State the estimated local loading matrices for cluster-specific factors, denote as $\widehat{\mathbf{\Gamma}}$ and $\widehat{\mathbf{\Lambda}}$, are finally given by the leading $k$ ($r$) eigenvectors of $\widehat{\mathbf{M}}^{*}_{1}$ and $\widehat{\mathbf{M}}^{*}_{2}$ respectively.
\end{algorithmic}
\end{algorithm}

\subsection{$K$-means clustering algorithm}\label{sec2.4}
In this section, we propose a $K$-means algorithm to identify the hidden cluster memberships along both the row and column dimensions of $\mathbf{X}_t$. We begin with the estimation of an upper bound for the number of clusters. Let \(\widehat{m}\) be the number of eigenvalues of \(|\widehat{\mathbf{\Gamma}}\widehat{\mathbf{\Gamma}}^{\top}|\) that are greater than \(1-\log^{-1}(T)\), where \(\widehat{m}\) serves as an upper bound of the number of row clusters. The upper bound for the number of column clusters, denoted by \(\widehat{n}\), can be obtained by counting the number of eigenvalues of \(|\widehat{\mathbf{\Lambda}}\widehat{\mathbf{\Lambda}}^{\top}|\) that are greater than \(1-\log^{-1}(T)\), i.e.,
\begin{equation}\label{mnhat}
    \widehat{m}=\sum_i\mathbb{I}\left\{\lambda_i(|\widehat{\mathbf{\Gamma}}\widehat{\mathbf{\Gamma}}^{\top}|)> 1-\log^{-1}(T)\right\},\ \ \widehat{n}=\sum_j\mathbb{I}\left\{\lambda_j(|\widehat{\mathbf{\Lambda}}\widehat{\mathbf{\Lambda}}^{\top}|)> 1-\log^{-1}(T)\right\},
\end{equation}
where $\mathbb{I}\{\cdot\}$ is an indicator function. Although \(\widehat{m}\) and \(\widehat{n}\) provide the upper bounds for $m$ and $n$, respectively (see Theorem \ref{groupnum2} in Section \ref{sec3}), our empirical experience shows that the estimators \(\widehat{m}\) and \(\widehat{n}\) equal the true number of clusters with high probability, see our simulation results in Table \ref{Table14} and Table \ref{Table24} in Section \ref{sec4}.  In what follows, we will illustrate the intuition behind the definition of $\widehat{m}$,  and that of $\widehat{n}$ can be understood in a similar way. Note that \({\mathbf{\Gamma}}{\mathbf{\Gamma}}^{\top}\) is a block diagonal matrix with $m$ blocks and all the non-zero eigenvalues equal to 1. Therefore the dominant eigenvalue for each of the latent $m$ blocks in \(\widehat{\mathbf{\Gamma}}\widehat{\mathbf{\Gamma}}^{\top}\) is greater than or at least very close to 1. Moreover, by Perron-Frobenius’s theorem, the largest eigenvalue of \(|\widehat{\mathbf{\Gamma}}_i\widehat{\mathbf{\Gamma}}_i^{\top}|\) is strictly greater than the other eigenvalues of  \(|\widehat{\mathbf{\Gamma}}_i\widehat{\mathbf{\Gamma}}_i^{\top}|\) under  Assumption \ref{ass2} (iv) in Section 2. Combining with the theoretical analysis, the above method to find the upper bound of the cluster number is feasible.
 
In the following, we define two similarity measure matrices, $\mathbf{D}$ and $\mathbf{K}$ based on ${\mathbf{\Gamma}}$ and ${\mathbf{\Lambda}}$, respectively. Specifically, let $\mathbf{D}=(d_{i,j})$ and $\mathbf{K}=(K_{i,j})$ be the $p\times p$ and $q\times q$ matrix with
\begin{equation}\label{d}
    d_{i,j}=|{\boldsymbol{\gamma}}^{\top}_i{\boldsymbol{\gamma}}_j|/({\boldsymbol{\gamma}}^{\top}_i{\boldsymbol{\gamma}}_i\cdot {\boldsymbol{\gamma}}^{\top}_j{\boldsymbol{\gamma}}_j)^{1/2},\ 1\leq i,j\leq p,
\end{equation}
\begin{equation}\label{k}
    K_{i,j}=|{\boldsymbol{\lambda}}^{\top}_i{\boldsymbol{\lambda}}_j|/({\boldsymbol{\lambda}}^{\top}_i{\boldsymbol{\lambda}}_i\cdot {\boldsymbol{\lambda}}^{\top}_j{\boldsymbol{\lambda}}_j)^{1/2},\ 1\leq i,j\leq q,
\end{equation}
and in practice, we define $\widehat{\mathbf{D}}$ as
\begin{equation}\label{dhat}
    \widehat{\mathbf{D}}=(\widehat{d}_{i,j})_{p\times p}=\left(|\widehat{\boldsymbol{\gamma}}^{\top}_i\widehat{\boldsymbol{\gamma}}_j|/({\widehat{\boldsymbol{\gamma}}}^{\top}_i\widehat{\boldsymbol{\gamma}}_i\cdot \widehat{\boldsymbol{\gamma}}^{\top}_j\widehat{\boldsymbol{\gamma}}_j)^{1/2}\right),\ 1\leq i,j\leq p,
\end{equation}
where $\widehat{\boldsymbol{\gamma}}^{\top}_i$ denotes the $i$-th row of matrix $\widehat{\mathbf{\Gamma}}$. We obtain $\widehat{\mathbf{K}}$ by performing similar steps based on $\mathbf{\widehat{\Lambda}}$, i.e., for the $i$-th row of matrix $\widehat{\mathbf{\Lambda}}$, denote as $\widehat{\boldsymbol{\lambda}}^{\top}_i$,
\begin{equation}\label{khat}
    \widehat{\mathbf{K}}=(\widehat{K}_{i,j})_{q\times q}=\left(|\widehat{\boldsymbol{\lambda}}^{\top}_i\widehat{\boldsymbol{\lambda}}_j|/({\widehat{\boldsymbol{\lambda}}}^{\top}_i\widehat{\boldsymbol{\lambda}}_i\cdot \widehat{\boldsymbol{\lambda}}^{\top}_j\widehat{\boldsymbol{\lambda}}_j)^{1/2}\right),\ 1\leq i,j\leq q.
\end{equation}
We then apply $K$-means clustering algorithm to the rows or columns of \(\widehat{\mathbf{D}}\) and  \(\widehat{\mathbf{K}}\), respectively, forming $\hat{m}$ and $\hat{n}$ clusters and  determining each cluster memberships simultaneously.

At last, we summarize the whole biclustering procedures for matrix-valued time series in Algorithm \ref{alg3}, also as a conclusion of this section.
\begin{algorithm}[h]
\caption{{Biclustering procedure for matrix-valued time series} \label{alg3}}
\KwIn{Matrix-valued time series $\{\mathbf{X}_t\}_{t\leq T}$, a positive integer $l_0$.}
\KwOut{Biclustering results for $\{\mathbf{X}_t\}_{t\leq T}$ including cluster numbers and each cluster memberships.}
\begin{algorithmic}[1]
\State{\textbf{Estimation for the number of factors:} \
given data matrix $\mathbf{X}_t$, for $l=1,\dots,l_0$, define $\widehat{\mathbf{M}}_{0,1}$ and $\widehat{\mathbf{M}}_{0,2}$ in (\ref{sigmahatM01}) and (\ref{M02hat}), obtain the estimators $(\widehat{k}_0,\widehat{k})$, $(\widehat{r}_0,\widehat{r})$ by (\ref{210}) and (\ref{facnum_r});}
\State{\textbf{Estimation for the global loading matrices of common factors:} \
obtain $\widehat{\mathbf{R}}$ and $\widehat{\mathbf{C}}$ by Algorithm \ref{alg1};}
\State{\textbf{Estimation for the loading matrices of cluster-specific factors:} \
obtain the estimators of local loading matrices $\widehat{\mathbf{\Gamma}}$ and $\widehat{\mathbf{\Lambda}}$ by Algorithm \ref{alg2};}
\State{\textbf{$K$-means clustering:} \
calculate the row (column) cluster numbers \(\widehat{m}\) (\(\widehat{n}\)) by (\ref{mnhat}), apply $K$-means clustering algorithm to form $\hat{m}$ and $\hat{n}$ clusters to the rows or columns of $\widehat{\mathbf{D}}$ and $\widehat{\mathbf{K}}$ defined in (\ref{dhat}) and (\ref{khat}) respectively.}
\end{algorithmic}
\end{algorithm}

\section{Asymptotic Properties}\label{sec3}
In Section \ref{sec2}, we propose our model setup along with its estimation details, and develop an algorithm for biclustering. The primary objective of this section is to study the asymptotic properties of the estimators under the setting that $T$, $p$ and $q$ grow to infinity. In addition to Assumptions \ref{ass1}-\ref{ass3} given in Section \ref{sec2}, the following regularity assumptions are needed to obtain the theoretical properties.
\begin{as1}\label{ass4}
We assume that the vector-valued process 
 $\{\boldsymbol{g}_t\}$ and $\{\boldsymbol{f}_t\}$ both have mean $\boldsymbol{0}$ and we set 

\begin{enumerate}
    \item[(i)] the process $\{\boldsymbol{g}_t\}$ and $\{\boldsymbol{f}_t\}$ are $\alpha$-mixing with mixing coefficients satisfying the conditions $(pq)^{-2}\sum_{k=1}^{\infty} \alpha_1(k)^{1-2 / \gamma}<\infty$ and $p^{-(2-2\delta_1)}q^{-(2-2\delta_2)}\sum_{k=1}^{\infty} \alpha_2(k)^{1-2 / \gamma}<\infty$ respectively, for some $\gamma>2$, where
	$$
	\alpha_s(k)=\sup _i \sup _{A \in {\mathcal{F}^s}_{-\infty}^i, B \in {\mathcal{F}^s}_{i+k}^{\infty}}|P(A \cap B)-P(A) P(B)|,\ \ s=1,2,
	$$
	where ${\mathcal{F}^1}_i^j$ and ${\mathcal{F}^2}_i^j$ are  the $\sigma$-field generated by $\left\{\bm{g}_t : i \leq t \leq j\right\}$ and $\left\{\bm{f}_t : i \leq t \leq j\right\}$, respectively.
    \item[(ii)] $\frac{1}{T}\sum_{t=1}^{T-l}\mathbb{E}\left|\boldsymbol{f}_{t+l,i}\boldsymbol{g}_{t,j}-\mathbb{E}\left(\boldsymbol{f}_{t+l,i}\boldsymbol{g}_{t,j}\right)\right|^2=O_p\left(p^{2-\delta_1}q^{2-\delta_2}T^{-1}\right)$ and 
    
    $\frac{1}{T}\sum_{t=1}^{T-l}\mathbb{E}\left|\boldsymbol{g}_{t+l,i}\boldsymbol{f}_{t,j}-\mathbb{E}\left(\boldsymbol{g}_{t+l,i}\boldsymbol{f}_{t,j}\right)\right|^2=O_p\left(p^{2-\delta_1}q^{2-\delta_2}T^{-1}\right)$ for $i,j=1,\dots,pq$.
\end{enumerate}
\end{as1}
\begin{as1}\label{ass5}
Let $\boldsymbol{e}_t^0=vec(\mathbf{E}_t^0)=\mathbf{A}\boldsymbol{\epsilon}_t$, where $\mathbf{A}$ is a $pq\times pq$ constant matrix with $\left\|\mathbf{A}\right\|$ bounded by a positive constant independent of $pq$. Furthermore, 
$\boldsymbol{\epsilon}_t$ is an $MA(\infty)$ process, i.e., $\boldsymbol{\epsilon}_t=\sum_{s=0}^{\infty}\phi_s\boldsymbol{\eta}_{t-s}$, where $\sum_{s=0}^{\infty}|\phi_s|<\infty,\boldsymbol{\eta}_t=(\eta_{t,1},\dots,\eta_{t,pq})^{\top}$, with $\eta_{t,i}$ being i.i.d. across $t$ and $i$ with mean 0, variance 1 and $E(\eta_{t,i}^4)<\infty$.
\end{as1}
Assumption \ref{ass4} is a standard assumption in the matrix factor model literature, see for example \cite{Wang2019,Chen2020ConstrainedFM,YU2022,Fan2023}. Similar as \cite{Chen2020ConstrainedFM} and \cite{YU2022}, we only require the factor process to satisfy the mixing condition in Assumption \ref{ass4}(i).
Assumption \ref{ass4}(ii) allows weak correlation between global (strong) and cluster-specific (weak) factors across time, rows and columns. Assumption \ref{ass5} defines a serial corelation structure for $\boldsymbol{e}_t^0$, which is similar to the Assumption 4 in \cite{Yao2023} for vector factor model and relaxes the  white idiosyncratic noise assumption  in \cite{Wang2019}.

Under Assumptions \ref{ass1}-\ref{ass5}, we first derive the following two theorems in terms of estimating the factor loading spaces. Firstly, Theorem \ref{th1} presents the asymptotic properties of $\widehat{\mathbf{R}}$ and $\widehat{\mathbf{C}}$ under the spectral norm.

\begin{thm1}\label{th1}
    Under Assumptions \ref{ass1}-\ref{ass5} with $T,\ p,\ q\to\infty$, $(k_0,r_0)$ fixed and given,
    we have \begin{equation}\label{thm1eq}
 \begin{aligned}
 \left\|\widehat{\mathbf{R}}\widehat{\mathbf{R}}^{\top}-\mathbf{R}\mathbf{R}^{\top}\right\|&=O_p\left(T^{-1 / 2}\right),\\
\left\|\widehat{\mathbf{C}}\widehat{\mathbf{C}}^{\top}-\mathbf{C}\mathbf{C}^{\top}\right\|&=O_p\left(T^{-1 / 2}\right).
 \end{aligned}
 \end{equation}
 \end{thm1}
 
Theorem \ref{th1} establishes the convergence rates of the estimated global factor loading spaces, from which we can see that the convergence rate depends on the horizon of the observed time series. It also shows that the estimation errors for $\widehat{\mathbf{R}}$ and $\widehat{\mathbf{C}}$ are asymptotically immune to the increase of $p$ and $q$. When $p$ and $q$ grow, the curse of dimensionality is offset by the information brought by new incoming series. From Theorem \ref{th1} we can see that the asymptotic convergence rates of the global loading space estimators are the same as those derived for the case when there is only strong factors discussed in \cite{Wang2019}.

In the following we proceed to present the convergence rates of the estimators $\widehat{\mathbf{\Gamma}}$ and $\widehat{\mathbf{\Lambda}}$ in Theorem \ref{th2} below.

 \begin{thm1}\label{th2}
Under Assumptions \ref{ass1}-\ref{ass5} and the condition $p^{\delta_1}q^{\delta_2}T^{-\frac{1}{2}}=o\left(1\right)$ as $T$, $p$ and $q$ tend to $\infty$ while $(k,r)$ are known and fixed, we have
    \begin{equation}\label{thm2eq}
 \begin{aligned}
\left\|\widehat{\mathbf{\Gamma}}\widehat{\mathbf{\Gamma}}^{\top}-\mathbf{P}_{\mathbf{R}\bot\mathbf{\Gamma}}\right\|&=O_p\left(p^{\delta_1}q^{\delta_2}T^{-\frac{1}{2}}\right),\\
\left\|\widehat{\mathbf{\Lambda}}\widehat{\mathbf{\Lambda}}^{\top}-\mathbf{P}_{\mathbf{C}\bot\mathbf{\Lambda}}\right\|&=O_p\left(p^{\delta_1}q^{\delta_2}T^{-\frac{1}{2}}\right).
 \end{aligned}
 \end{equation}
  \end{thm1}

Theorem \ref{th2} shows that the estimators $\widehat{\mathbf{\Gamma}}$ and $\widehat{\mathbf{\Lambda}}$ are consistent under some mild conditions. Indeed, $p^{\delta_1}q^{\delta_2}T^{-1/2}=o(1)$ is a scaling condition to derive the consistency of the estimators for loading spaces in matrix factor models \citep{Wang2019,liu2022identification,Gao2021ATF}. It shows that as long as $T^{1/2}$ increases faster than $p^{\delta_1}q^{\delta_2}$ asymptotically, $\widehat{\mathbf{\Gamma}}$ and $\widehat{\mathbf{\Lambda}}$ converge to $\mathbf{P}_{\mathbf{R}\bot\mathbf{\Gamma}}$ and $\mathbf{P}_{\mathbf{R}\bot\mathbf{\Gamma}}$, respectively. When $\delta_i=0,i=1,2$, i.e., all the factors are strong or pervasive, we can achieve the standard convergence rate $T^{-1}$.

Equipped with $\widehat{\mathbf{\Gamma}}$ and $\widehat{\mathbf{\Lambda}}$, we obtain the estimators of the  upper bounds for the number of clusters, i.e., \(\widehat{m}\) and \(\widehat{n}\) as described in Section \ref{sec2.4}. The following theorem provides the theoretical guarantee for such estimators.

\begin{thm1}\label{groupnum2}
Under Assumption \ref{ass1}-\ref{ass5} and the condition $p^{\delta_1}q^{\delta_2}\cdot \max\{k^{1/2},r^{1/2}\}\cdot \log T=o(T^{1/2})$, we have $P(\widehat{m}\geq m)\rightarrow 1$ and $P(\widehat{n}\geq n)\rightarrow 1$, as $T,p,q\rightarrow \infty$.
\end{thm1}
Theorem \ref{groupnum2} shows that the probability of underestimating the cluster numbers tends to zero. When $k$ and $r$ increase slowly with $p$, $q$, we need an additional condition $p^{\delta_1}q^{\delta_2}\cdot \max\{k^{1/2},r^{1/2}\}\cdot \log T=o(T^{1/2})$ to obtain the consistency, which is slightly stronger than condition $p^{\delta_1}q^{\delta_2}T^{-1}=o(1)$ in Theorem \ref{th2}.

In the following we assume that $m$ and $n$ are known and investigate the theoretical error rate in term  of clustering. Clustering along the row dimension is used for illustration here, and similar theoretical analysis applies parallelly to column dimension. We define a set $\mathfrak{O}_{d}$ which consists of all $p \times p$ matrices with $m$ distinct rows. Let 
\begin{equation}
    \mathbf{D}_0=\arg\min_{\mathbf{O} \in \mathfrak{O}_{d}}\|{\mathbf{D}}-\mathbf{O}\|_{F}^{2},
\end{equation} 
where $\mathbf{D}$ is defined in (\ref{d}). For any $p$-dimensional vector $\boldsymbol{g}$ with its elements taking integer values ranging from 1 to $m$, let $\mathfrak{O}_{d}(\boldsymbol{g})=\{\mathbf{O}\in \mathfrak{O}_{d}$: {two rows of} $\mathbf{O}$ {are the same if and only if the corresponding two elements of} $\boldsymbol{g}$ {are the same}$\}$. As pointed out in \cite{Yao2023}, the $m$ distinct rows of $\mathbf{D}_0$ would be the centers of the $m$ clusters identified by the $K$-means method applied to the rows of $\mathbf{D}$. As $\mathbf{D}$ is unknown, we identify the $m$ clusters based on its estimator $\widehat{\mathbf{D}}$ defined in (\ref{dhat}). Denote $\boldsymbol{g}_0$ as a $p$-dimensional vector whose first $p_1$ elements are 1 and the next  $p_2$ elements are 2, and so on so forth, with $m$ being the value of the last $p_m$ elements. Note that, given the block diagonal structure of $\mathbf{\Gamma}$, we can conclude that $\mathbf{D}_0\in \mathfrak{O}_{d}(\boldsymbol{g}_0)$.

To obtain a misclustering error that asymptotically converge to zero, we need an additional assumption illustrated as follows, which is similar to Assumption 6 in \cite{Yao2023}. Let $\mathbf{K}_0=\arg\min_{\mathbf{O} \in \mathfrak{O}_{d}}\|{\mathbf{K}}-\mathbf{O}\|_{F}^{2}$, where $\mathbf{K}$ is defined in (\ref{k}).

\begin{as1}\label{groupass6}
For some constant $c>0$,
$$
\begin{aligned}
    \min _{\mathbf{O} \in \mathfrak{O}_{d}(\boldsymbol{g})}\|{\mathbf{D}}-\mathbf{O}\|_{F}^{2}&\geq  \|{\mathbf{D}}-\mathbf{D}_0\|_{F}^{2}+c\tau(\boldsymbol{g})p^{1-\delta_1},\\
    \min _{\mathbf{O} \in \mathfrak{O}_{d}(\widetilde{\boldsymbol{g}})}\|{\mathbf{K}}-\mathbf{O}\|_{F}^{2}&\geq  \|{\mathbf{K}}-\mathbf{K}_0\|_{F}^{2}+c\tau(\widetilde{\boldsymbol{g}})q^{1-\delta_2},
\end{aligned}
$$
 $\tau(\boldsymbol{g})$ ($\tau(\dot{\boldsymbol{g}})$) denotes the number of misclassified components by
partition $\boldsymbol{g}$ ($\widetilde{\boldsymbol{g}}$), where for any $p$ ($q$)-vector $\boldsymbol{g}$ ($\widetilde{\boldsymbol{g}}$) with its elements taking integer values ranging from 1 to $m$ ($n$).
\end{as1}
The objective of $K$-means clustering algorithm is to minimize the sum of distances between data points and their assigned clusters. Data points that are nearest to a centroid are clustered together within the same category. Assumption \ref{groupass6} implies that $\|{\mathbf{D}}-\mathbf{O}\|_{F}^{2}$ will increase when the number of misplaced members of partition $\boldsymbol{g}$ becomes large, which is necessary for the $K$-means clustering algorithm. Together with Assumption \ref{groupass6}, the following theorem provides a convergence rate of the misclustering error which characterize the performance of Algorithm \ref{alg3}.

\begin{thm1}\label{thm5}
Let Assumption \ref{ass1}-\ref{groupass6} hold and the number of row clusters $m$ and column clusters $n$ are assumed to be known. Denoted by $\widehat{\tau}_1$ ($\widehat{\tau}_2$) the number of misclassified components in $\widehat{\mathbf{D}}$ ($\widehat{\mathbf{K}}$) by the $K$-means clustering. Then as $T,p,q\rightarrow\infty$,
\begin{equation}\label{35}
    \widehat{\tau}_1/p=O_p(p^{-\delta_1/2}),\ \widehat{\tau}_2/q=O_p(q^{-\delta_2/2}).
\end{equation}
\end{thm1}
Theorem \ref{thm5} implies that the misclassification rates of the $K$-means method converge to zero, which guarantees the accuracy of the clustering results.

The above analysis is based on that the number of common and cluster-specific factors $k_0$, $r_0$, $k$ and $r$ are known in advance. In practice, both the factors and loadings are unobservable and the numbers of factors need to be estimated, which is required as a preliminary step in order to use our methodology. In the following, we establish the theoretical guarantee for the eigenvalue-ratio based estimators proposed in Section \ref{sec2.1}. To this end, we need an additional assumption on eigenvalues.

\begin{as1}\label{factornum}
\(\mathbf{M}_{0,1}\) and \(\mathbf{M}_{0,2}\) have $k+k_0$ and $r+r_0$ distinct positive eigenvalues, respectively, while \(\mathbf{M}^*_{0,1}\) and \(\mathbf{M}^*_{0,2}\) have $k$ and $r$ distinct positive eigenvalues, respectively.
\end{as1}

Assumption \ref{factornum} is standard in large factor models and has been commonly made in the related literature. The nonzero eigenvalues of \(\mathbf{M}_{0,1}\), \(\mathbf{M}_{0,2}\), \(\mathbf{M}^*_{0,1}\) and \(\mathbf{M}^*_{0,2}\) are assumed to be distinct from each other for  further identifiability issue, and we refer, for example, to \cite{Wang2019,liu2022identification,Fan2023} for similar assumptions. With Assumption \ref{factornum}, in Theorem \ref{facnumthm} below we establish that the one-pass estimators of the number of factors are consistent.

\begin{thm1}\label{facnumthm}
   Under Assumptions \ref{ass1}-\ref{factornum} with $p^{\delta_1}q^{\delta_2}T^{-\frac{1}{2}}=o\left(1\right)$, when $p,q,T\to\infty$, we have $\mathbb{P}\left(\widehat{k}_0=k_0\right)\to 1$, $\mathbb{P}\left(\widehat{k}=k\right)\to 1$, $\mathbb{P}\left(\widehat{r}_0=r_0\right)\to 1$, and $\mathbb{P}\left(\widehat{r}=r\right)\to 1$.
\end{thm1}

Theorem \ref{facnumthm} specifies the asymptotic behavior for the ratios of the cumulated eigenvalues used in estimating the numbers of factors in Section \ref{sec2.1}. It implies that the estimators converge to the true ones in probability, which provides a theoretical underpinning for the estimators in (\ref{210}) and (\ref{facnum_r}).

By Theorem \ref{th1}-\ref{facnumthm}, we thoroughly established the theoretical guarantees of our proposed methodology in Section \ref{sec2} and in the next section we will further verify the asymptotic results by simulation studies.

\section{Simulation Studies}\label{sec4}
In this section, we evaluate the finite sample performances of Algorithm \ref{alg1}-\ref{alg3}, using synthetic simulated data. For all settings, the reported results are based on 500 replications and $l=1,\dots,5$.

We begin with describing the generation mechanism of the observed data matrices according to model (\ref{model}). In detail, the elements of the matrices $\mathbf{R}$, $\mathbf{C}$, $\mathbf{\Gamma}_i$, and $\mathbf{\Lambda}_j$ are independently sampled from a uniform distribution $U(-1,1)$, with $i = 1, \dots, m$ and $j = 1, \dots, n$. The sequences  $\{\bm{g}_t\}$ and $\{\boldsymbol{f}_t\}$ are assumed to follow independent first-order autoregressive (AR(1)) and moving average (MA(1)) processes, respectively, with Gaussian innovations. Similarly, the components of the noise vector $\boldsymbol{e}_t^0$ are modeled as independent MA(1) processes with Gaussian innovations distributed as $N(0, 0.25)$. The autoregressive and moving average coefficients are randomly drawn from the union of two disjoint intervals, $U{(-0.95, -0.4) \cup (0.4, 0.95)}$, ensuring sufficient variability while avoiding near-unit-root behavior. Furthermore, the standard deviations of the components of $\boldsymbol{g}_t$ and $\boldsymbol{f}_t$ are independently sampled from the uniform distribution $U(1,2)$. These settings are designed to thoroughly evaluate the robustness and effectiveness of our proposed method under diverse stochastic structures.

In order to evaluate the method we proposed, we set different values for $T$, $p$, and $q$ so as to verify the asymptotic properties established in Section \ref{sec3}. We consider the following two scenarios:
\textbf{Scenario I:}  $T=400,\ m=3\ \text{and } n=3,$
\textbf{Scenario II:} $T=500,\ m=5\ \text{and } n=4,$
with $k_0=k_1=\dots=k_m=3,\ p_1=\dots=p_m,$
$r_0=r_1=\dots=r_n=2,\ q_1=\dots=q_n$, hence $k=m\cdot k_0$, $r=n\cdot r_0$, $p=m\cdot p_1$ and $q=n\cdot q_1$. In Section \ref{sec4.1}, we first evaluate the finite sample performance of the one-pass estimators for the numbers of factors proposed in Section \ref{sec2.1}. Then we evaluate the finite sample performance of the estimators for the common and cluster-specific loading matrices in Section \ref{sec4.2}. At last, we 
evaluate the biclustering accuracy in Section \ref{sec4.3}.

\subsection{Estimation of  the numbers of factors}\label{sec4.1}
\begin{table}[ht!]
		\centering
		\caption{The relative frequencies of $\widehat{k}_0=k_0$, $\widehat{k}_0+\widehat{r}_0=k_0+r_0$, $\widehat{k}=k$ and $\widehat{k}+\widehat{r}=k+r$ for \textbf{Scenario I} with 500 replications.}
		\label{Table11}
			\begin{tabular}{cccccc} 
				\toprule 
    \multirow[b]{2}{*}{Factor number results}&\multirow[b]{2}{*}{$l_0$}&   \multicolumn{4}{c}{$(p_1,q_1)$} \\
				\cmidrule(lr){3-6}
&&$(10,10)$&$(10,15)$&$(20,20)$&$(25,20)$ \\
				\hline
$\widehat{k}_0=k_0$
&1&.678&.680&.862&.854\\
&2&.622&.612&.830&.822\\
&3&.598&.608&.760&.786\\
&4&.532&.542&.718&.722\\
&5&.502&.554&.670&.690\\
$\widehat{k}_0+\widehat{r}_0=k_0+r_0$
&1&.888&.878&1&1\\
&2&.810&.856&.998&1\\
&3&.858&.866&.996&1\\
&4&.800&.818&1&1\\
&5&.848&.848&.998&1\\
$\widehat{k}=k$
&1&.886&.992&.940&.954\\
&2&.902&.906&.916&.960\\
&3&.862&.910&.940&.906\\
&4&.826&.866&.920&.910\\
&5&.818&.872&.930&.856\\
$\widehat{k}+\widehat{r}=k+r$
&1&.980&1&1&1\\
&2&.982&1&1&1\\
&3&.990&1&1&1\\
&4&.986&1&1&1\\
&5&.984&1&1&1\\
			\bottomrule 
			\end{tabular}
	\end{table}
 \begin{table}[ht!]
		\centering
		\caption{The relative frequencies of $\widehat{k}_0=k_0$, $\widehat{k}_0+\widehat{r}_0=k_0+r_0$, $\widehat{k}=k$ and $\widehat{k}+\widehat{r}=k+r$ for \textbf{Scenario II} with 500 replications.}
		\label{Table21}
			\begin{tabular}{cccccc} 
				\toprule 
    \multirow[b]{2}{*}{Factor number results}&\multirow[b]{2}{*}{$l_0$}&   \multicolumn{4}{c}{$(p_1,q_1)$} \\
				\cmidrule(lr){3-6}
&&$(10,10)$&$(10,15)$&$(20,20)$&$(25,20)$ \\
				\hline
$\widehat{k}_0=k_0$
&1&.794&.818&.902&.904\\
&2&.734&.750&.900&.858\\
&3&.698&.750&.822&.876\\
&4&.666&.712&.788&.810\\
&5&.622&.686&.780&.796\\
$\widehat{k}_0+\widehat{r}_0=k_0+r_0$
&1&.806&.842&1&1\\
&2&.772&.762&.992&1\\
&3&.768&.810&.998&1\\
&4&.760&.810&.998&1\\
&5&.770&.764&1&1\\
$\widehat{k}=k$
&1&.976&.974&.962&.964\\
&2&.974&.954&.948&.956\\
&3&.938&.962&.942&.942\\
&4&.916&.942&.910&.908\\
&5&.894&.932&.906&.906\\
$\widehat{k}+\widehat{r}=k+r$
&1&.998&1&1&1\\
&2&.994&.998&1&1\\
&3&.992&1&1&1\\
&4&.990&1&1&1\\
&5&.990&1&1&1\\
			\bottomrule 
			\end{tabular}
	\end{table}
Accurate estimation of the number of factors plays a pivotal role in the estimation of matrix factor model and subsequent clustering analysis. We study the empirical performance of the proposed estimators for the factor numbers defined  in (\ref{210}) and (\ref{facnum_r}). Table \ref{Table11} and Table \ref{Table21} present the frequencies of exact estimation over 500 replications under Scenario I and Scenario II, from which we can conclude that our proposed methodology, based on $\widehat{R}_{i,j},i=1,2$ defined in Section \ref{sec2.1}, can determine the numbers of both strong and weak factors and is not sensitive to the choice of $l_0$. As the sample size $T$ or dimension size $(p,q)$ increases, the frequencies that $\widehat{k}_0=k_0$, $\widehat{k}_0+\widehat{r}_0=k_0+r_0$, $\widehat{k}=k$ and $\widehat{k}+\widehat{r}=k+r$ tends to 1 as well, which also matches our theoretical results. 

\subsection{Estimation  of the factor loading spaces}\label{sec4.2}
In this section, we study the estimation accuracy of the estimators for loading matrices and report the estimation errors for the factor loading spaces in Table \ref{Table13}-\ref{Table23}. We adopt a measure to quantify the distance of two linear spaces as in \cite{Wang2019,YU2022,Fan2023,He2024Matrix}. Let $\mathbf{S}_i,i=1,2$
 be full-rank matrices in $\mathbb{R}^{p\times q_i}$. Let ${\mathbf{O}}_i$ be the matrix whose columns form an orthonormal basis of $\mathcal{M}(\mathbf{S}_i)$ for $i=1,2$, then the distance between column spaces of $\mathcal{M}(\mathbf{S}_1)$ and $\mathcal{M}(\mathbf{S}_2)$ can be measured by
\begin{equation}
\mathcal{D}\left({\mathbf{S}}_1,\mathbf{S}_2\right)=\left(1-\frac{1}{\min\{q_1,q_2\}}\text{tr}({\mathbf{O}}_1{\mathbf{O}}_1^{\top}{\mathbf{O}}_2{\mathbf{O}}_2^{\top})\right)^{1/2}.
\end{equation}
It can be inferred that $\mathcal{D}\left({\mathbf{S}}_1,\mathbf{S}_2\right)$ is a measure that ranges between 0 and 1. The value of $\mathcal{D}\left({\mathbf{S}}_1,\mathbf{S}_2\right)$ equals 0 if $\mathcal{M}({\mathbf{S}}_1)\subseteq\mathcal{M}({\mathbf{S}}_2)$ or $\mathcal{M}({\mathbf{S}}_2)\subseteq\mathcal{M}({\mathbf{S}}_1)$, and 1 if and only if $\mathcal{M}({\mathbf{S}}_1)\perp\mathcal{M}({\mathbf{S}}_2)$. 

Detailed estimation errors are presented in  Table \ref{Table13} and Table \ref{Table23} which contain the means and standard deviations (in parentheses) of $\mathcal{D}(\widehat{\mathbf{R}},\mathbf{R})$, $\mathcal{D}(\widehat{\mathbf{C}},\mathbf{C})$, $\mathcal{D}\left(\widehat{\mathbf{\Gamma}},(\mathbf{I}_p-{\mathbf{R}}{\mathbf{R}}^{\top})\cdot\mathbf{\Gamma}\right)$ and $\mathcal{D}\left(\widehat{\mathbf{\Lambda}},(\mathbf{I}_p-{\mathbf{C}}{\mathbf{C}}^{\top})\cdot\mathbf{\Lambda}\right)$ from Algorithm \ref{alg3} over 500 replications. In the following simulations the numbers of factors are given as priori. As shown in Table \ref{Table13} and Table \ref{Table23}, the estimation errors (also the standard errors) tend to decrease as $T$ increases, as well as  $(p,q)$ increases, though less pronounced. It can also be concluded that the proposed estimation method for the loading spaces is not sensitive to the choice of time lag $l_0$.
\begin{table}[ht!]
		\centering
		\caption{Averaged estimation errors and standard errors (in parentheses) of $\mathcal{D}\left(\widehat{\mathbf{R}},\mathbf{R}\right)$, $\mathcal{D}\left(\widehat{\mathbf{C}},\mathbf{C}\right)$, $\mathcal{D}\left(\widehat{\mathbf{\Gamma}},(\mathbf{I}_p-{\mathbf{R}}{\mathbf{R}}^{\top})\cdot\mathbf{\Gamma}\right)$ and $\mathcal{D}\left(\widehat{\mathbf{\Lambda}},(\mathbf{I}_p-{\mathbf{C}}{\mathbf{C}}^{\top})\cdot\mathbf{\Lambda}\right)$ for \textbf{Scenario I} with 500 replications. Both the row factor numbers and column factor numbers are assumed to be known.}
		\label{Table13}
			\begin{tabular}{cccccc} 
				\toprule 
    \multirow[b]{2}{*}{Estimation errors }&\multirow[b]{2}{*}{$l_0$}&   \multicolumn{4}{c}{$(p_1,q_1)$} \\
				\cmidrule(lr){3-6}
&&$(10,10)$&$(10,15)$&$(20,20)$&$(25,20)$ \\
				\hline
$\mathcal{D}\left(\widehat{\mathbf{R}},\mathbf{R}\right)$
&1&.031(.011)&.025(.008)&.031(.009)&.021(.007)\\
&2&.032(.011)&.026(.009)&.031(.010)&.022(.007)\\
&3&.033(.011)&.027(.009)&.033(.011)&.023(.007)\\
&4&.034(.012)&.027(.009)&.034(.011)&.023(.007)\\
&5&.035(.012)&.028(.013)&.034(.011)&.024(.007)\\
$\mathcal{D}\left(\widehat{\mathbf{C}},\mathbf{C}\right)$
&1&.024(.008)&.025(.007)&.024(.007)&.015(.005)\\
&2&.025(.009)&.026(.008)&.025(.007)&.015(.005)\\
&3&.026(.009)&.026(.008)&.026(.008)&.016(.005)\\
&4&.027(.009)&.026(.008)&.027(.009)&.016(.005)\\
&5&.027(.011)&.027(.012)&.027(.009)&.017(.005)\\
$\mathcal{D}\left(\widehat{\mathbf{\Gamma}},(\mathbf{I}_p-{\mathbf{R}}{\mathbf{R}}^{\top})\cdot\mathbf{\Gamma}\right)$
&1&.039(.007)&.031(.005)&.032(.004)&.024(.002)\\
&2&.036(.007)&.029(.005)&.030(.004)&.023(.003)\\
&3&.034(.006)&.028(.005)&.029(.004)&.022(.003)\\
&4&.033(.006)&.027(.004)&.028(.005)&.021(.003)\\
&5&.033(.006)&.026(.006)&.028(.005)&.022(.003)\\
$\mathcal{D}\left(\widehat{\mathbf{\Lambda}},(\mathbf{I}_p-{\mathbf{C}}{\mathbf{C}}^{\top})\cdot\mathbf{\Lambda}\right)$
&1&.030(.005)&.029(.003)&.025(.003)&.017(.002)\\
&2&.029(.004)&.028(.004)&.024(.003)&.016(.002)\\
&3&.027(.004)&.026(.003)&.023(.003)&.016(.002)\\
&4&.026(.005)&.026(.004)&.023(.004)&.015(.002)\\
&5&.026(.005)&.025(.006)&.022(.004)&.015(.002)\\
				\bottomrule
			\end{tabular}
	\end{table}
\begin{table}[ht!]
		\centering
		\caption{Averaged estimation errors and standard errors (in parentheses) of $\mathcal{D}\left(\widehat{\mathbf{R}},\mathbf{R}\right)$, $\mathcal{D}\left(\widehat{\mathbf{C}},\mathbf{C}\right)$, $\mathcal{D}\left(\widehat{\mathbf{\Gamma}},(\mathbf{I}_p-{\mathbf{R}}{\mathbf{R}}^{\top})\cdot\mathbf{\Gamma}\right)$ and $\mathcal{D}\left(\widehat{\mathbf{\Lambda}},(\mathbf{I}_p-{\mathbf{C}}{\mathbf{C}}^{\top})\cdot\mathbf{\Lambda}\right)$ for \textbf{Scenario II} with 500 replications. Both the row factor numbers and column factor numbers are assumed to be known.}
		\label{Table23}
			\begin{tabular}{cccccc} 
				\toprule 
    \multirow[b]{2}{*}{Estimation errors}&\multirow[b]{2}{*}{$l_0$}&   \multicolumn{4}{c}{$(p_1,q_1)$} \\
				\cmidrule(lr){3-6}
&&$(10,10)$&$(10,15)$&$(20,20)$&$(25,20)$ \\
				\hline
$\mathcal{D}\left(\widehat{\mathbf{R}},\mathbf{R}\right)$
&1&.026(.007)&.021(.006)&.026(.006)&.018(.005)\\
&2&.026(.007)&.022(.006)&.027(.007)&.027(.007)\\
&3&.028(.007)&.023(.006)&.028(.007)&.028(.007)\\
&4&.028(.007)&.023(.006)&.028(.007)&.029(.007)\\
&5&.029(.007)&.023(.006)&.029(.007)&.029(.008)\\
$\mathcal{D}\left(\widehat{\mathbf{C}},\mathbf{C}\right)$
&1&.018(.005)&.018(.005)&.018(.005)&.011(.003)\\
&2&.018(.005)&.018(.005)&.018(.005)&.017(.005)\\
&3&.020(.005)&.019(.005)&.020(.005)&.017(.005)\\
&4&.020(.005)&.020(.005)&.019(.005)&.018(.005)\\
&5&.020(.005)&.020(.005)&.020(.005)&.018(.005)\\
$\mathcal{D}\left(\widehat{\mathbf{\Gamma}},(\mathbf{I}_p-{\mathbf{R}}{\mathbf{R}}^{\top})\cdot\mathbf{\Gamma}\right)$
&1&.034(.004)&.027(.003)&.027(.002)&.021(.001)\\
&2&.031(.004)&.025(.003)&.025(.002)&.025(.002)\\
&3&.029(.003)&.023(.003)&.024(.002)&.024(.002)\\
&4&.027(.003)&.022(.002)&.023(.002)&.023(.002)\\
&5&.026(.003)&.021(.002)&.021(.002)&.023(.002)\\
$\mathcal{D}\left(\widehat{\mathbf{\Lambda}},(\mathbf{I}_p-{\mathbf{C}}{\mathbf{C}}^{\top})\cdot\mathbf{\Lambda}\right)$
&1&.023(.002)&.021(.002)&.018(.001)&.013(.001)\\
&2&.020(.002)&.020(.002)&.017(.002)&.015(.001)\\
&3&.019(.002)&.018(.002)&.016(.002)&.014(.002)\\
&4&.018(.002)&.018(.002)&.016(.002)&.014(.002)\\
&5&.018(.002)&.017(.002)&.015(.002)&.014(.002)\\
				\bottomrule 
			\end{tabular}
	\end{table}


\subsection{Clustering accuracy}\label{sec4.3}
 \begin{table}[ht!]
		\centering
		\caption{The means and standard deviations (in parentheses) of the estimated cluster number and cluster accuracy for \textbf{Scenario I} with 500 replications, where the numbers of factors $r_0$, $r$, $k_0$ and $k$ are known.}
		\label{Table14}
			\begin{tabular}{cccccc} 
				\toprule 
    \multirow[b]{2}{*}{Clustering results}&\multirow[b]{2}{*}{$l_0$}&   \multicolumn{4}{c}{$(p_1,q_1)$} \\
				\cmidrule(lr){3-6}
&&$(10,10)$&$(10,15)$&$(20,20)$&$(25,20)$ \\
				\hline
$\widehat{m}$
&1&3.044(.224)&3.064(.261)&3(0)&3(0)\\
&2&3.062(.250)&3.042(.211)&3(0)&3(0)\\
&3&3.056(.247)&3.056(.239)&3.002(.045)&3(0)\\
&4&3.058(.242)&3.070(.263)&3(0)&3(0)\\
&5&3.054(.226)&3.050(.227)&3(0)&3(0)\\
$\widehat{n}$
&1&2.970(.203)&3(0)&3(0)&3(0)\\
&2&2.964(.216)&3(0)&3(0)&3(0)\\
&3&2.950(.244)&3(0)&3.002(.045)&3(0)\\
&4&2.972(.208)&2.998(.045)&3(0)&3(0)\\
&5&2.950(.244)&2.998(.045)&3(0)&3(0)\\
$clusteracc-m$
&1&.955(.079)&.951(.086)&.998(.006)&.998(.005)\\
&2&.953(.079)&.957(.077)&.997(.007)&.998(.005)\\
&3&.956(.071)&.952(.084)&.998(.009)&.998(.005)\\
&4&.951(.081)&.952(.081)&.997(.009)&.998(.005)\\
&5&.959(.068)&.953(.078)&.997(.007)&.998(.005)\\
$clusteracc-n$
&1&.964(.056)&.984(.023)&.991(.015)&.990(.016)\\
&2&.963(.052)&.984(.027)&.991(.014)&.992(.014)\\
&3&.967(.051)&.984(.026)&.990(.017)&.989(.020)\\
&4&.964(.056)&.984(.024)&.990(.019)&.989(.015)\\
&5&.962(.056)&.985(.027)&.989(.016)&.990(.016)\\
			\bottomrule 
			\end{tabular}
	\end{table}
\begin{table}[ht!]
		\centering
		\caption{The means and standard deviations (in parentheses) of the estimated cluster number and cluster accuracy  for \textbf{Scenario II} with 500 replications, where the numbers of factors $r_0$, $r$, $k_0$ and $k$ given.}
		\label{Table24}
			\begin{tabular}{cccccc} 
				\toprule 
    \multirow[b]{2}{*}{Cluster results}&\multirow[b]{2}{*}{$l_0$}&   \multicolumn{4}{c}{$(p_1,q_1)$} \\
				\cmidrule(lr){3-6}
&&$(10,10)$&$(10,15)$&$(20,20)$&$(25,20)$ \\
				\hline
$\widehat{m}$
&1&5.160(.388)&5.140(.359)&5(0)&5(0)\\
&2&5.144(.357)&5.146(.359)&5(0)&5(0) \\
&3&5.154(.372)&5.152(.365)&5(0)&5(0)\\
&4&5.180(.395)&5.144(.357)&5.006(.077)&5(0)\\
&5&5.154(.367)&5.112(.316)&5(0)&5(0)\\
$\widehat{n}$
&1&4.008(.126)&4(0)&4(0)&4(0)\\
&2&4.008(.109)&4(0)&4(0)&4.002(.045) \\
&3&4.006(.077)&4(0)&4(0)&4(0)\\
&4&3.996(.110)&4(0)&4(0)&4(0)\\
&5&4.006(.134)&4(0)&4(0)&4.002(.045)\\
$clusteracc-m$
&1&.975(.053)&.979(.046)&.999(.004)&.999(.002)\\
&2&.981(.043)&.981(.043)&.999(.003)&.999(.002)\\
&3&.978(.049)&.978(,053)&.999(.003)&.999(.002)\\
&4&.976(.049)&.981(.045)&.999(.007)&1(0)\\
&5&.981(.038)&.985(.037)&.999(.005)&.999(.002)\\
$clusteracc-n$
&1&.978(.036)&.991(.014)&.993(.010)&.994(.010)\\
&2&.978(.035)&.990(.018)&.993(.012)&.994(.012)\\
&3&.980(.035)&.991(.015)&.993(.012)&.994(.012)\\
&4&.980(.035)&.984(.018)&.994(.008)&.993(.011)\\
&5&.977(.041)&.990(.016)&.993(.011)&.993(.010)\\
			\bottomrule 
			\end{tabular}
	\end{table}
In this section, we present the results for the estimated number of clusters and clustering accuracy based on 500 replications, given the numbers of factors $r_0$, $r$, $k_0$, and $k$. Recall the $K$-means clustering algorithm in Section \ref{sec2.4}, where $\widehat{m}$ and $\widehat{n}$ are the estimated  upper bounds for $m$ and $n$, respectively. Given $k_i = 3$ and $r_j = 2$ for $i = 0, 1, \dots, m$ and $j = 0, 1, \dots, n$, it is worth noting that $\widehat{m} = m$ and $\widehat{n} = n$ almost always hold in our simulations (see Table \ref{Table14} and Table \ref{Table24}). The $\widehat{m}$ clusters with detailed cluster memberships are then obtained by applying $K$-means clustering algorithm to the rows of $\widehat{\mathbf{D}}$. Similarly, the rows of $\widehat{\mathbf{K}}$ are divided into $\widehat{n}$ clusters using the same $K$-means clustering algorithm. We calculate the number of correctly clustered elements and only report the simulation results for $\widehat{m}=m$ and $\widehat{n}=n$. Table \ref{Table14} and Table \ref{Table24} report the means and standard deviations of the correct specified rates across 500 replications, which clearly demonstrate that our biclustering method identifies the latent row/column clusters with a high degree of accuracy and the proportion of exact specification by $K$-means algorithm goes toward one as the dimension size $(p,q)$ grows.

\section{Real Data Analysis: Multinational Macroeconomic Indices}\label{sec5}
\subsection{Data Description and Prepocessing}
In this section, we apply our biclustring procedure to a multi-national macroeconomic indices dataset which is also ever studied in \cite{YU2022} and \cite{Fan2023}. The dataset is collected from Organizationfor Economic Co-operation and Development (OECD). It contains 10 macroeconomic indices across 8 countries over 130 quarters from 1988-Q1 to 2020-Q2. As shown in Section \ref{sec1}, the countries, which include the United States, the United Kingdom, Canada, France, Germany, Norway, Australia and New Zealand, can be roughly characterized as European, North American and Oceanian factors,  following their geographical partitions by common sense. The indices can be roughly divided into  4 major clusters, namely Consumer Price, Interest Rate, Production, and International Trade by common sense. In the following we begin to analyze the $8\times 10$ matrix-valued time series and investigate whether our unsupervised biclustering method would result in the same clusters as from common sense. 

We first use a similar logarithmic transformation and differencing operations as in \cite{Fan2023}. We further standardize each of the transformed series to avoid the effects of non-zero mean or diversified variances. By implementing the proposed method, we accomplish various empirical objectives, including the estimation of factor models, the unsupervised biclustering and a rolling validation prediction. Subsequent sections showcase the detailed results for this multi-national macroeconomic indices dataset.

\subsection{The numbers of factors and row/column clusters}
\begin{figure}[ht!]
    \centering
    \clearpage
\includegraphics[width=10cm,height=6cm]{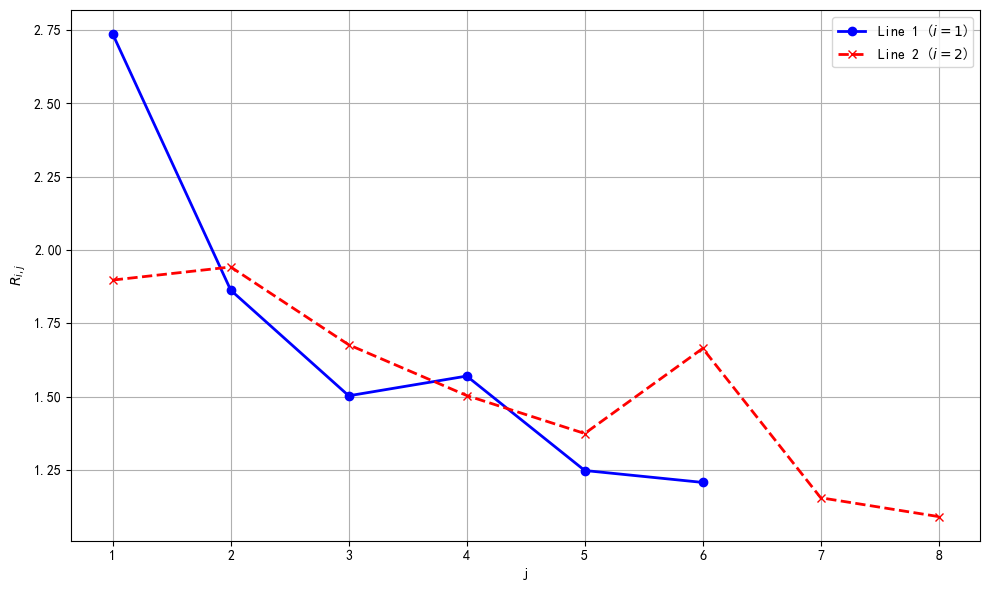}
    \caption{Plot of the $\widehat{R}_{1,j}$ (blue) and $\widehat{R}_{2,j}$ (red) to estimate the numbers of strong and weak factors, the points marked on lines are the locations where local maximum values occur.}
    \label{figure1}
\end{figure}

Following the procedures  in Algorithm \ref{alg3}, the first step is to determine the numbers of row and column factors. In this empirical study we fixed $l_0=5$, as the simulation results show that our procedure is not sensitive to the choice of $l_0$. Figure \ref{figure1} illustrates the eigenvalue-ratios $\widehat{R}_{i,j},i=1,2$ calculated according to (\ref{29}) and (\ref{facnum_r}).
Specifically, it shows that $\widehat{R}_{1,1}$ is much larger than all the others for row factors. By (\ref{210}), we choose $\widehat{k}_0=1$ and $\widehat{k}_0+\widehat{k}=4$ which is reasonable. Figure \ref{figure1} also plots the estimated number of strong and weak column factors, and it indicates that $\widehat{R}_{2,j}$ are the 1st and the 2nd largest local maximum value with $j=6$ and $j=2$ respectively. Hence we take $\widehat{r}_0=2$ and $\widehat{r}_0+\widehat{r}=6$. By Figure \ref{figure1}, it
suggests that there are 4 weak column factors.

\begin{figure}[ht!]
    \centering
    \clearpage
\includegraphics[width=15cm,height=6cm]{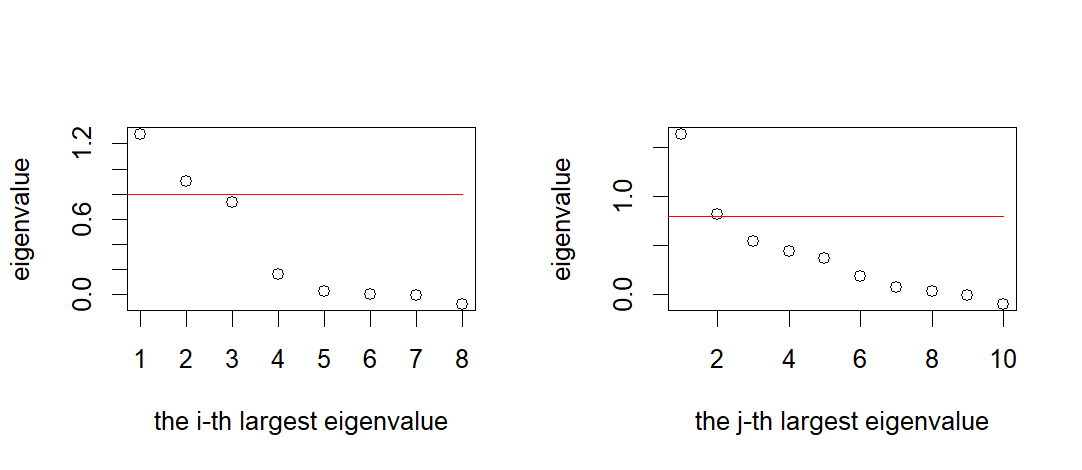}
    \caption{The i-th largest eigenvalues of $\left|\widehat{\mathbf{\Gamma}}\widehat{\mathbf{\Gamma}}^{\top}\right|$ (left) and the j-th largest eigenvalues of $\left|\widehat{\mathbf{\Lambda}}\widehat{\mathbf{\Lambda}}^{\top}\right|$ (right) when $\widehat{k}_0=1$,  $\widehat{k}_0+\widehat{k}=4$, $\widehat{r}_0=2$ and $\widehat{r}_0+\widehat{r}=6$, the red line is $1-\log^{-1}T$.}
    \label{figure3}
\end{figure}

Following the procedures in Algorithm \ref{alg3}, we then obtain the estimators $\widehat{\mathbf{\Gamma}}$ and $\widehat{\mathbf{\Lambda}}$, and perform the $K$-means clustering algorithm for the rows of matrix $\widehat{\mathbf{D}}$ and $\widehat{\mathbf{K}}$ defined in (\ref{dhat}) and (\ref{khat}), respectively. {By Figure \ref{figure3} which plots the eigenvalues of $\left|\widehat{\mathbf{\Gamma}}\widehat{\mathbf{\Gamma}}^{\top}\right|$ and $\left|\widehat{\mathbf{\Lambda}}\widehat{\mathbf{\Lambda}}^{\top}\right|$, we obtain the estimators of the number of row/column clusters as  $\widehat{m}=2$ and $\widehat{n}=2$. However, taking into the clusters by common sense as comparison, we also consider $\widehat{m}=3$ and $\widehat{n}=3,4$ in the subsequent analysis.}

\subsection{Clustering results}
\begin{figure}[ht!]
    \centering
    \clearpage
\includegraphics[width=15cm,height=5cm]{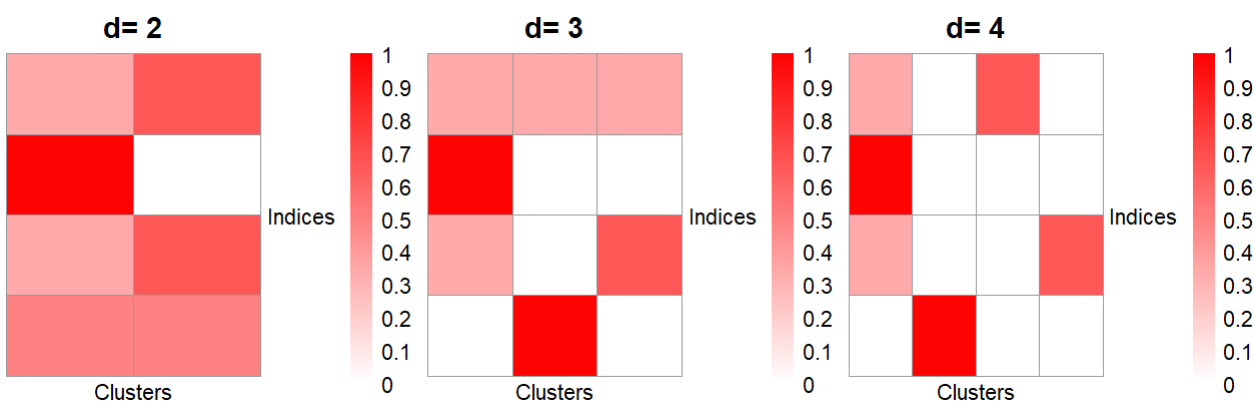}
    \caption{Heat-maps of the distributions of the indices in each of the 4 groups (corresponding to 4 rows) over $\widehat{n}$ clusters (corresponding to $\widehat{n}$ columns), with $\widehat{n}=2,3$ and 4.}
    \label{figure5}
\end{figure}

We analyze the biclustering results   in this section. We mainly focus on the clustering results along the column direction, i.e, the macroeconomic indices, which is of greater interest for this real dataset. To present the identified $\widehat{n}$ clusters, we define a $4\times \widehat{n}$ matrix with $\widehat{n}_{ij}/\widehat{n}_i$ as its $(i,j)$-th element, where $\widehat{n}_i$ is the number of the indices in the i-th sector, and $\widehat{n}_{ij}$ is the number of the indices in the i-th cluster which are allocated in the j-th cluster. Thus $\widehat{n}_{ij}/\widehat{n}_i\in [0,1]$ and $\sum_j \widehat{n}_{ij}/\widehat{n}_i=1$. The heat-maps of this $4\times \widehat{n}$ matrix for $\widehat{n}=2,3,4$ are presented in Figure \ref{figure5}. For $\widehat{n}=2$, Cluster 1 mainly contains the indicator in Interest Rate and Cluster 2 mainly contains the indices in Consumer Price and Production. When $\widehat{n}$ increases to 3, the most obvious change is that the indicator International Trade is merged into the second category. When we choose $\widehat{n}=4$, Cluster 1 still mainly contains the indicator in Interest Rate, Cluster 2 consists of International Trade, Cluster 3 and Cluster 4 contain the indices in Consumer Price and Production, respectively. When $\widehat{n}$ is increased from 2 to 4, Interest Rate indices are always clustered into a single group. Additionally, there is always an indicator in both Consumer Price and Production that is categorized with the Interest Rate.
\subsection{Rolling validation}
 In this section, we use a rolling validation procedure similar as in \cite{YU2022} to further compare the performances of our proposed method with those of the existing state-of-the-art competitors  including:
\begin{itemize}
    \item \textbf{Auto-Cross-Correlation Estimation (ACCE) method: } for matrix factor model (\ref{factor_model}), estimate the dynamic signal part of $\mathbf{X}_t$ by the method in \cite{Wang2019};
    \item \textbf{$\alpha$-PCA method: } for matrix factor model (\ref{factor_model}), use an estimation method proposed by \cite{Fan2023} to estimate $\mathbf{R}$, $\mathbf{C}$ and $\mathbf{F}_t$ with parameter $\alpha=0$;
    \item \textbf{Projected Estimation (PE) method: } for matrix factor model (\ref{factor_model}), obtain the estimators of loading matrices $\mathbf{R}$, $\mathbf{C}$ and factor matrix $\mathbf{F}_t$ by the Algorithm 1 in \cite{YU2022}.
\end{itemize}

In view of the small sample size in this example, for each quarter $t$ from 2008-Q1 to 2020-Q2, we repeatedly use the observations before $t$ to estimate the matrix- factor model. The estimated loadings are then used to calculate the mean squared error at time point $t$. Specifically, let $\mathbf{X}_t$ and $\widehat{\mathbf{X}}_t$ be the observed and estimated matrix-valued time series in the $t$-th quarter, and define
\begin{equation}\label{mse}
    \overline{\text{MSE}}=\frac{1}{T\times p\times q}\sum_{t=1}^{T}\left\|\widehat{\mathbf{X}}_t-{\mathbf{X}}_t\right\|^2_F,\ 
    \text{where}\ T=130,p=8,q=10.
\end{equation}
We try different factor numbers in a range of reasonable values and report the corresponding averaged MSE in Table \ref{table_real}.

From Table \ref{table_real}, we can conclude that the mean squared error for the ACCE method is the largest, the $\alpha$-PCA and PE method in \cite{Fan2023} and \cite{YU2022} are comparable but both are  inferior to our method in most cases. In contrast to the matrix factor model in (\ref{factor_model}), our new model (\ref{model}) takes into account the influence of both strong and weak factors, and the effect of the incorporation of the cluster-specific factors is further reflected by the different performance of various methods in  terms of $\overline{\text{MSE}}$.
 It is clear that the estimation accuracy is further improved by incorporating additional weak factors  information extracted from the error terms. Indeed, weak factors may contribute less to the explanatory power of the model, but the presence of weak factors would increase the complexity of the model, and in this real analysis further enhances the estimation accuracy. 
\begin{table}[ht!]
		\centering
		\caption{The averaged MSE defined in (\ref{mse}) by different methods.}
		\label{table_real}
			\begin{tabular}{cccc|cccc} 
				\toprule 
    \multirow[b]{2}{*}{$k_0$}&\multirow[b]{2}{*}{$k$}&\multirow[b]{2}{*}{$r_0$}&\multirow[b]{2}{*}{$r$}&\multicolumn{4}{c}{$\overline{\text{MSE}}$}\\
    &&&&ACCE&$\alpha$-PCA ($\alpha=0$) &PE&Our method\\
    \hline
    1&1&2&2&0.994&0.956&0.952&\textbf{0.930}\\
    1&1&3&3&0.994&0.956&0.952&\textbf{0.875}\\
    1&1&4&4&0.994&0.956&0.952&\textbf{0.795}\\
    2&2&2&2&0.827&0.727&0.749&0.820\\
    2&2&3&3&0.827&0.727&0.749&0.762\\
    2&2&4&4&0.827&0.727&0.749&\textbf{0.713}\\
    3&3&2&2&0.645&0.604&0.594&0.629\\
    3&3&3&3&0.645&0.604&0.594&\textbf{0.590}\\
    3&3&4&4&0.645&0.604&0.594&\textbf{0.543}\\
			\bottomrule 
			\end{tabular}
	\end{table}

\section{Conclusion}\label{sec6}
In this study, we developed a novel unsupervised learning methodology for biclustering high dimensional matrix-valued time series, grounded in a latent two-way factor structure. Each cluster is characterized by its unique row- and column-specific factors, in addition to shared matrix factors that influence all matrix time series. The proposed approach accounts for the dynamic dependence of the factors, and a two-step projection estimation procedure is proposed for the entirely new factor model with both strong/common and weak/cluster-specific factors. We rigorously established the asymptotic properties for the estimators of loading matrices. Moreover, a eigenvalue-ratio based method is introduced to determine the numbers of factors, and we demonstrate the consistency of the estimators with explicit convergence rates for the common factors, cluster-specific factors, and latent clusters. Numerical illustration with both simulated data as well as a real data example is reported to demonstrate the empirical usefulness and effectiveness of our proposed method. 

Future research directions include extending the proposed framework to multi-mode clustering for tensor-valued time series, which is more challenging  and warrants further investigation.

\section*{Acknowledgements}
This work is supported by NSF China (12171282) and Qilu Young Scholars Program of Shandong University, China. The authors are grateful to the organizer Catherine C. Liu, the attendants  Dan Yang (HKU), Long Yu (SUFE), etc., at "The 1st Workshop of Data Science: Factor-Analysis Learning --- Inference, Bayesian, \& Practices", for their helpful discussions. Indeed, this work is inspired by the first author Yong He as a discussant of professor Qiwei Yao's talk "Factor Modelling for Clustering High-dimensional Time Series" at the workshop.

\bibliographystyle{model2-names}
\bibliography{proposal2}

\end{document}